\documentclass[12pt]{spieman}  
\usepackage{amsmath}  
\usepackage[]{graphicx}
\usepackage{setspace}
\usepackage{tocloft}

\title{Shaped Pupil Lyot Coronagraphs: High-Contrast Solutions for Restricted Focal Planes} 

\author{Neil T. Zimmerman,\supscr{a} A J Eldorado Riggs,\supscr{a}, N. Jeremy Kasdin\supscr{a}, Alexis Carlotti\supscr{b}, Robert J. Vanderbei\supscr{c}}

\affiliation{\supscrsm{a}Princeton University, Department of Mechanical and Aerospace Engineering, Engineering Quadrangle, Princeton, NJ 08544, USA\\
\supscrsm{b}Institut de Plan\'{e}tologie et d'Astrophysique, Centre National de la Recherche Scientifique, F-38000 Grenoble, France\\
\supscrsm{c}Princeton University, Department of Operations Research and Financial Engineering, Sherrerd Hall, Princeton, NJ 08544, USA
}

\cftpagenumbersoff{figure}
\cftpagenumbersoff{table} 
\begin{document} 

\newcommand{\pasp}{PASP}
\newcommand{\pasj}{PASJ}
\newcommand{\aj}{AJ}
\newcommand{\apjl}{ApJ}
\newcommand{\apj}{ApJ}
\newcommand{\apjs}{ApJS}
\newcommand{\apss}{Ap\&SS}
\newcommand{\araa}{ARA\&A}
\newcommand{\aapr}{A\&ARv}
\newcommand{\aap}{A\&A}
\newcommand{\aaps}{A\&AS}
\newcommand{\nat}{Nature}
\newcommand{\procspie}{Proc.~SPIE}
\newcommand{\jrasc}{JRASC}
\newcommand{\memras}{MmRAS}
\newcommand{\mnras}{MNRAS}
\newcommand{\na}{New A}
\newcommand{\nar}{New A Rev.}
\newcommand{\ao}{Applied Opt.}

\maketitle 

\begin{abstract}

Coronagraphs of the \textit{apodized pupil} and \textit{shaped pupil} varieties use the Fraunhofer diffraction properties of amplitude masks to create regions of high contrast in the vicinity of a target star. Here we present a hybrid coronagraph architecture in which a binary, hard-edged shaped pupil mask replaces the gray, smooth apodizer of the \textit{apodized pupil Lyot coronagraph} (APLC). For any contrast and bandwidth goal in this configuration, as long as the prescribed region of contrast is restricted to a finite area in the image, a shaped pupil is the apodizer with the highest transmission. We relate the starlight cancellation mechanism to that of the conventional APLC. We introduce a new class of solutions in which the amplitude profile of the Lyot stop, instead of being fixed as a padded replica of the telescope aperture, is jointly optimized with the apodizer. Finally, we describe \textit{shaped pupil Lyot coronagraph} (SPLC) designs for the baseline architecture of the WFIRST-AFTA coronagraph. These SPLCs help to enable two scientific objectives of the WFIRST-AFTA mission: (i) broadband spectroscopy to characterize exoplanet atmospheres in reflected starlight and (ii) debris disk imaging.

\end{abstract}


{\noindent \footnotesize{\bf Address all correspondence to}: Neil T. Zimmerman, Space Telescope Science Institute, 3700 San Martin Drive, Baltimore, MD 21218, USA; E-mail:  \linkable{ntz@stsci.edu} }

\begin{spacing}{1}   

\section{Introduction}
\label{sect:intro}  

The last two decades have witnessed tremendous advances in high-contrast imaging technology, in tandem with the emergence of exoplanet research. There is now a mature and growing assortment of instrument concepts devised to isolate the light of an exoplanet from its host star and acquire its spectrum. Stellar coronagraphs descended from Bernard Lyot's invention represent a major component of this effort, complementing and intersecting the innovations in interferometry, adaptive optics, wavefront control, and data processing. With these tools in place, several exoplanet imaging programs at large ground-based observatories are underway\cite{Hinkley2011PASP,Macintosh2014PNAS,Beuzit2008SPIE}. Their observations have led to discoveries and astrophysical measurements that are steering theories of planet formation, planetary system evolution, and planet atmospheres\cite{Oppenheimer2013ApJ,Macintosh2015Sci,Millar-Blanchaer2015ApJ,Quanz2015ApJ}. Meanwhile, laboratory testbeds are setting the stage for yet more ambitious instruments on new space telescopes\cite{Trauger2007Natur,Lawson2013SPIE,Mazoyer2014AA,NDiaye2015SPIE}.

A coronagraph alters the point spread function of a telescope so that a region of the image normally dominated by starlight is darkened by destructive interference, enabling observations of faint surrounding structures and companions. Starlight cancellation is accomplished with a group of optical elements that operate on the complex field of the propagating beam. The classical Lyot coronagraph functions with a pair of simple masks: one an opaque occulting spot at the focus, and second a ``Lyot stop'' to block the outer edge of the re-collimated on-axis beam before it is re-imaged\cite{Lyot1939MNRAS}. To take advantage of the diffraction-limited imaging capabilities of high-order adaptive optics (AO) systems, beginning in the 1990s classical Lyot designs were revised for high-contrast stellar coronagraphy\cite{Golimowski1992, Nakajima1994, Malbet1996, Mouillet1997, Sivaramakrishnan2001, Oppenheimer2004}. Through Fourier optical analysis and modeling, researchers soon discovered the remarkable performance benefits of apodizing the entrance pupil of a coronagraph\cite{Guyon2000SPIE, Aime2002, Soummer2003circ}. Since then, the transmission profile of this apodizer mask has been a topic of vigorous study\cite{Soummer2005APLCfATA1, Aime2005PASP, Abe2008, Soummer2009APLCfATA2, Soummer2011APLCfATA3, NDiaye2015APLCfATA4}. One resulting family of designs, the \textit{apodized pupil Lyot coronagraph} (APLC), has been successfully integrated with several AO-fed cameras to facilitate deep observations of young exoplanetary systems at near-infrared wavelengths.

As an alternative to a coronagraph with two or more mask planes, pupil apodization by itself is perhaps the simplest and oldest way to reject unwanted starlight from a telescope image\cite{Herschel1847,Barnard1909}. Fraunhofer diffraction theory dictates the way any change in the shape or transmission profile of the entrance pupil redistributes a star's energy in the image plane. This relationship can be used to design an apodizer whose point spread function (PSF) has a zone of high contrast near the star, without additional coronagraph masks. This is the \textit{shaped pupil} approach developed by N. J. Kasdin and collaborators, who pioneered the optimization of apodizers with binary-valued transmission\cite{Spergel2001, Kasdin2003, Vanderbei2003, Kasdin2005}. In recent years, shaped pupil solutions have evolved to work around arbitrary two-dimensional telescope apertures, in parallel with similar breakthroughs in APLC design\cite{Enya2010, Carlotti2011,Vanderbei2012,Carlotti2012,Carlotti2013AFTA,Haze2015}. The relative simplicity of a single mask, however, comes with a sacrifice in how close the dark search region can be pushed toward the star. At the contrast levels relevant to exoplanet imaging, the smallest feasible shaped pupil inner working angle is between 3--4 $\lambda/D$\cite{Kasdin2005}.

Shaped pupil coronagraphs (SPCs) and Lyot coronagraphs (both classical and APLC) both rely on masks that operate strictly on the transmitted amplitude of the propagating beam. Numerous coronagraph designs have been introduced that incorporate phase masks\cite{Roddier1997PASP,Rouan2000PASP,Mawet2009OExpr}, and pupil remapping via aspheric mirrors and/or static deformations\cite{Guyon2005PIAA,Pueyo2013ACAD}. In general, coronagraphs that manipulate phase in addition to amplitude can achieve higher performance in terms of inner working angle and throughput than SPCs and APLCs. For example, the vector vortex coronagraph has a theoretical inner working limit of ~$\lambda/D$ separation from a star\cite{Mawet2005ApJ}. Recent theoretical innovations have improved the compatibility of phase-mask and pupil-remapping coronagraph concepts with segmented and obstructed telescope apertures\cite{Mawet2013ApJS,Carlotti2014AA,Guyon2014ApJ,Ruane2015SPIE}. For broad comparisons between coronagraph design families, see Refs.~\citenum{Lawson2013SPIE}, \citenum{Guyon2006ApJS}, \citenum{Guyon2009AIPC}, and~\citenum{Levine2009}.

Until the past two years, all SPC testbed experiments with wavefront control used free-standing shaped pupil designs with connected obstruction patterns. In particular, experiments in Princeton's High Contrast Imaging Laboratory (HCIL)\cite{Kasdin2005SPIE} and the High Contrast Imaging Testbed (HCIT)\cite{Lowman2004SPIE} at the Jet Propulsion Laboratory (JPL) have used ripple-style SPC masks along with two deformable mirrors in series.\cite{groff2013filtering, Riggs2013SPIE, riggs2015ekf}. In this issue, Cady et al.~ \cite{Cady2015JATIS} report the first experimental results with a non-freestanding SPC design. This mask was fabricated on a silicon wafer substrate with aluminized reflective regions and highly absorptive black silicon regions; the fabrication process is described in detail in this issue by K. Balasubramanian et al.~ \cite{Bala2015JATIS}. E. Cady et al.\ used a single deformable mirror in their experiments to create a single-sided dark hole from $4.4-11~\lambda/D$ in a $52$-degree wedge\cite{Cady2015JATIS}. They achieved $5.9\times10^{-9}$ contrast in a 2\% bandwidth about 550 nm and $9.1\times10^{-8}$ contrast in a 10\% bandwidth about 550 nm.

The Science Definition Team of NASA's Wide-Field InfraRed Survey Telescope - Astrophysics Focused Telescope Assets mission (WFIRST-AFTA) has proposed including a Coronagraph Instrument (CGI) to observe super-Earth and gas-giant exoplanets in reflected starlight at visible wavelengths\cite{Spergel2015WFIRST}. Coronagraph designs for WFIRST-AFTA must be compatible with its heavily obscured telescope aperture, broad filter passbands, and rapid development timeline. The SPC, recognized to match these demands, was selected as one of two baseline coronagraph technologies to undergo extensive testing at JPL in advance of the mission formulation\cite{Poberezhskiy}. The method under development in parallel with the SPC is the \textit{hybrid Lyot coronagraph} (HLC)\cite{Trauger2015JATIS,Seo2015JATIS}. The HLC departs from the classical Lyot approach by using a focal plane mask with a complex transmission profile\cite{Trauger2012SPIE}. The SPC and HLC can share the same optical path and wavefront control system. A third coronagraph type, the Phase-Induced Amplitude Apodization Complex Mask Coronagraph, is being pursued in parallel as a backup option\cite{Guyon2012SPIE, Guyon2014ApJ, Pluzhnik2015JATIS, Kern2015JATIS}.

In the course of our efforts to improve the SPC designs already meeting the minimum performance goals, we investigate a hybrid coronagraph architecture in which a binary shaped pupil functions as the apodizer mask in an APLC-like configuration\cite{Riggs2014SP, NDiaye2015AAS}. In effect, this expands on the idea first put forward by Cady et al.~\cite{Cady2009ApJ}, who designed a hard-edged, star-shaped apodizer for the Gemini Planet Imager's APLC. We identify this design category as the \textit{shaped pupil Lyot coronagraph} (SPLC). The SPLC offers a persuasive union of the virtues of the SPC and the APLC: a binary apodizer with achromatic transmission properties and promising fabrication avenues\cite{Bala2006SPIE,Enya2012,Bala2015JATIS}, and the relatively small inner working angle and robustness to aberrations of an APLC\cite{Sivaramakrishnan2008ApJ}.

\section{Lyot coronagraphy with an unobscured circular aperture}
\label{sec:circ_splc}

Although coronagraph designs for obscured apertures are the ones of the highest practical interest and relevance to WFIRST-AFTA and the general community, a clear circular telescope aperture offers a natural starting point to understand how the SPLC relates functionally to the conventional APLC. For one, circular symmetry simplifies the analytical formulation, the numerical optimization problem, and the interpretation. The same qualitative relationships that occur for a simple aperture will reappear for more complicated cases (for example, SP apodizer feature size and outer working angle). Furthermore, the clear circular aperture allows us to probe the ultimate limitations of pure amplitude Lyot coronagraphy, offering useful insights for exoplanet imaging mission design studies such as the recent Exo-C\cite{Stapelfeldt2015ExoC}.

\begin{figure}[htb]
\begin{center}
\begin{tabular}{c}
\includegraphics[height=4cm]{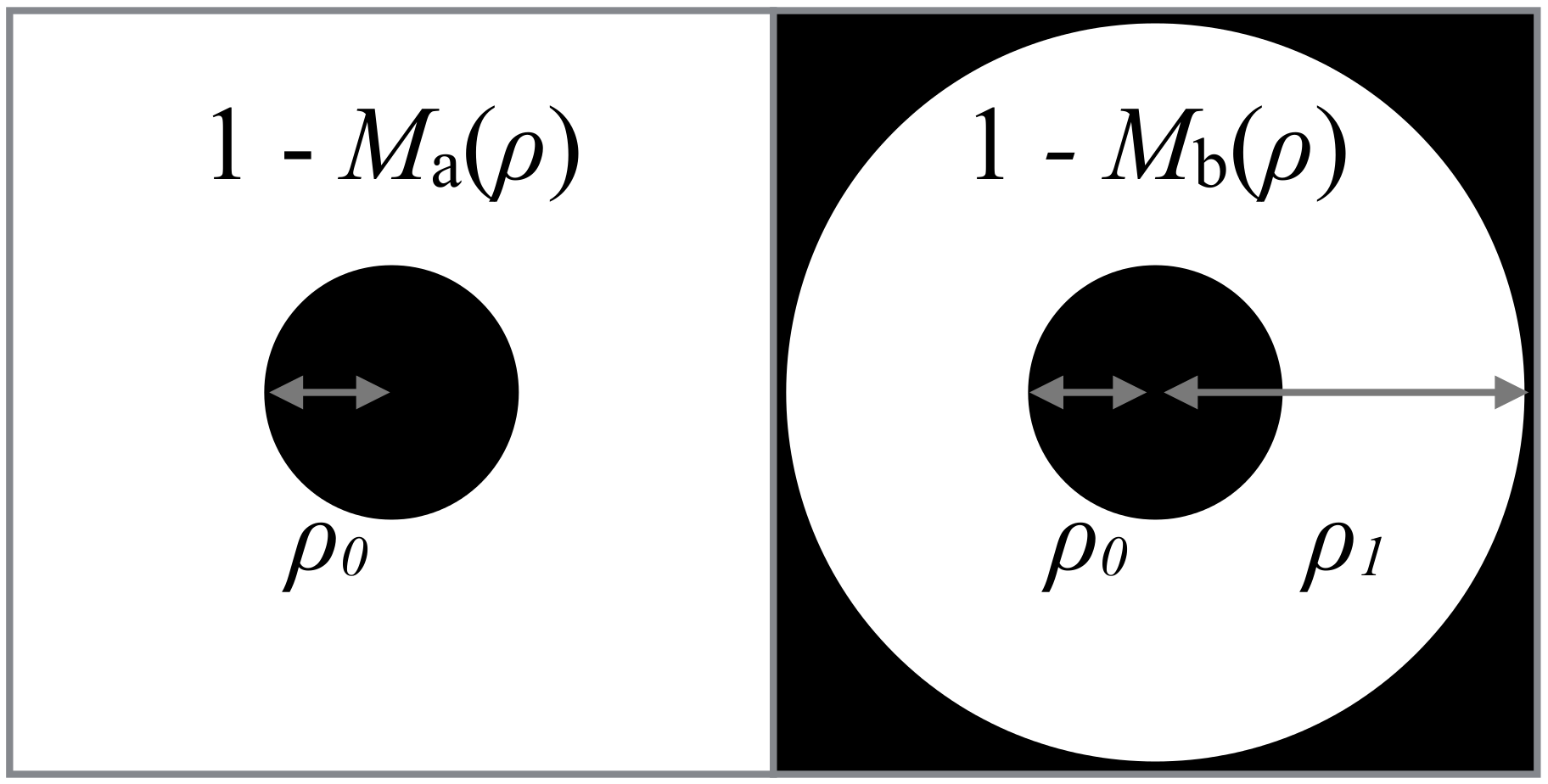}
\end{tabular}
\end{center}
\caption
{ \label{fig:FPM_diagram}
The two types of focal plane mask considered for the circular aperture Lyot coronagraph: occulting spot (left) and annular diaphragm (right). } 
\end{figure}

For each of our numerical SPLC experiments, we consider two forms of the focal plane mask (FPM), illustrated in Figure~\ref{fig:FPM_diagram}. First, the occulting spot of the conventional APLC, with radius $\rho_0$; second, an annular diaphragm with inner radius $\rho_0$ and outer radius $\rho_1$. In our descriptions of the on-axis field propagation, we will make use of the complement of the FPM transmission function. For the spot and diaphragm FPM cases, we label these $M_{\rm a}$ and $M_{\rm b}$, respectively. In terms of the radial spatial coordinate, $\rho$, they are defined as


\begin{equation}
\begin{aligned}
M_{\rm a} (\rho)  & = \Pi(\rho/(2\rho_0)) \\
& \mathrm{and}\\
M_{\rm b} (\rho) & = 1 - \Pi(\rho/(2\rho_1)) + \Pi(\rho/(2\rho_0)).\\
\label{eqn:circ_fpm_def}
\end{aligned}
\end{equation}

\noindent Here $\Pi\left(x\right)$ is the rectangle function, equal to unity inside $|x| < 1/2$ and zero elsewhere.

\begin{figure}[htb]
\begin{center}
\begin{tabular}{c}
\includegraphics[height=8cm]{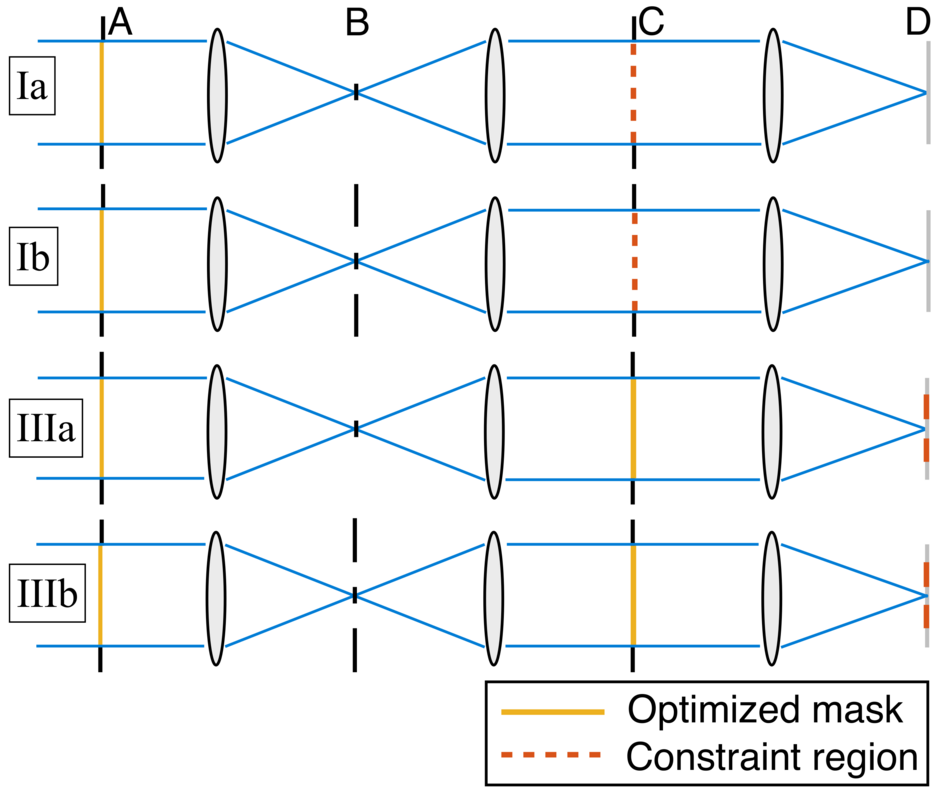}
\end{tabular}
\end{center}
\caption
{ \label{fig:circ_splc_diagrams}
Diagrams of a representative subset of the optimization schemes applied to the circular aperture Lyot coronagraph. From left to right, the critical mask planes are the apodizer (A), the focal plane mask (B), the Lyot stop (C), and the final image (D). The red (dashed) regions are where the on-axis field is constrained. The gold (solid) segments mark the masks that are optimized as free variables.}
\end{figure}

To find apodizer solutions for the circular SPLC, we use the same numerical optimization tools previously applied to shaped pupil mask designs. In addition to the two types of FPM above, we consider two different planes of field cancellation constraints, as diagrammed in Figure~\ref{fig:circ_splc_diagrams}. The results of all the circular SPLC trials are later summarized in Table~\ref{tab:circ_splc_results}. Details about our optimization method, including discrete algebraic models for the on-axis field propagation, and definitions of the linear program objectives and constraints, are given in Appendix~\ref{subsec:circsplc_appendix}.

\subsection{Focal occulting spot}
\label{sec:config1a}

Loosely following the nomenclature that the authors R. Soummer et al.\ formulated in Ref.~\citenum{Soummer2003circ}, we represent the scalar electric field in the entrance pupil, focal plane, and Lyot plane respectively by $\Psi_A$, $\Psi_B$, and $\Psi_C$. In a slight departure, we define the focal plane radial coordinate $\rho$ in units of image resolution elements ($f\lambda/D$) and the radial coordinate in the two conjugate pupil planes as $r$, normalized to the aperture diameter, $D$. For brevity, we implicitly apply the pupil cutoff function in all instances of $\Psi_A$ and $\Psi_C$, and we set $D$ to 1. These provisions allows us to succinctly express the on-axis scalar electric field in the Lyot plane after the occulting spot, in accordance with the Babinet principle:

\begin{equation}
\begin{aligned}
\Psi_{C}(r) & = \Psi_{A}(r) - \widehat{\left\{ \hat{\Psi}_{A}(\rho) M_{\rm a}(\rho) \right\}} \\
                  & = \Psi_{A}(r) - \left(2\pi\right)^2 \int\limits_{\rho=0}^{\rho_0} \int\limits_{r'=0}^{1/2} \Psi_{A}\left(r'\right) J_0\left(2\pi \rho r' \right) r' \mathrm{d}r' J_0\left(2\pi r \rho \right) \rho~\mathrm{d}\rho. \\
\end{aligned}
\label{eqn:aplc_Psi_C}
\end{equation}

\noindent Hatted variables denote the application of the Hankel transform, and $J_0$ is the zero-order Bessel function of the first kind.

Setting the condition for total field cancellation, $\Psi_C(r)=0$, leads to an integral equation of the variable function $\Psi_A(r)$. In Ref.~\citenum{Soummer2003circ} the authors Soummer et al.\ showed that the approximate solutions are a subset of Slepian's circular prolate spheroidal wave functions, originally published four decades prior\cite{Slepian1964}. The zero-order prolate spheroidal wave functions possess two exceptional apodization properties. First, they are by definition invariant to the finite Fourier transform, so the scalar field in the focal plane after the apodizer is equal to the unrestricted prolate function itself, to within a scale factor. Second, the prolate apodizer maximizes the concentration of energy in the focal plane, within a radius set by the eigenvalue $0 < \Lambda < 1$ of the integral equation\cite{Slepian1965}. Therefore, once a focal plane spot radius $\rho_0$ has been chosen, the apodizer for optimum monochromatic extinction $\Psi_A(r) = \Phi_{\Lambda}(r)$ is fully determined. Invoking the finite Hankel invariance property of $\Phi_{\Lambda}(r)$, one can start from Equation~\ref{eqn:aplc_Psi_C} and arrive at a simple expression for the residual on-axis Lyot plane electric field:

\begin{equation}
\begin{aligned}
\Psi_{C}(r) &= \Phi_{\Lambda}(r) - \Lambda \Phi_{\Lambda}(r) \\
                  &= \Phi_{\Lambda}(r) \left(1 - \Lambda \right) \\
\end{aligned}
\label{eqn:prolate_cancellation}
\end{equation}

\noindent Unlike an alternate configuration in which the prolate apodizer is combined with a Roddier $\pi$ phase-shifter\cite{Guyon2000SPIE}, for the opaque occulting spot the monochromatic on-axis cancellation is never complete because no prolate solution corresponding to $\Lambda = 1$ exists\cite{Soummer2003circ}. But $\Lambda$ is already 0.999 at $\rho_0 = 1.87~\lambda/D$, for example, and it can be made arbitrarily close to unity by further widening the occulting spot at the expense of the inner working angle.

As a first experiment, we start with the coronagraph model portrayed at the top of Figure~\ref{fig:circ_splc_diagrams}, which we label Config.\ Ia. For the clear circular aperture, an occulting focal plane spot of radius $\rho_0$, and a Lyot stop with the same diameter as the aperture, we ask what entrance apodizer results in a monochromatic Lyot field cancellation factor $1 - \Lambda$ while maximizing the overall field transmission. Our aim is to independently recover one of the canonical circular prolate apodizers presented in Ref.~\citenum{Soummer2003circ}, corresponding to $\Lambda = 0.999$ and $\rho = 1.87$. We form a linear program relating the discretized apodizer vector to the resulting Lyot field, using a Riemann sum representation of the Hankel transforms on the right hand side of Equation~\ref{eqn:aplc_Psi_C}. For details about this procedure, see Appendix~\ref{subsec:circsplc_appendix}.

\begin{figure}[htb]
\begin{center}
\begin{tabular}{c}
\includegraphics[width=0.9\textwidth]{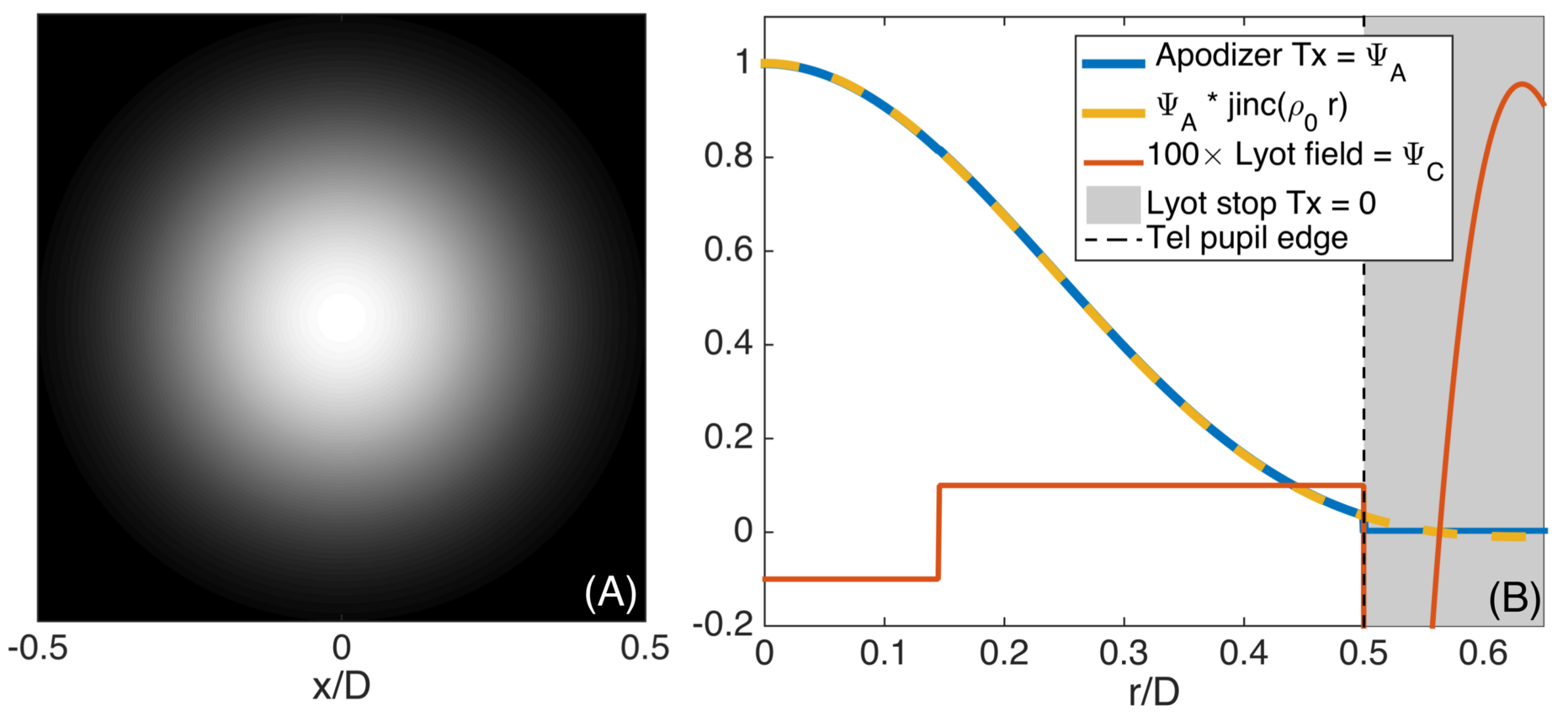}
\end{tabular}
\end{center}
\caption
{\label{fig:config1a} 
Circular Lyot coronagraph, Config.\ Ia: The apodizer (A) is optimized for maximum transmission while achieving a monochromatic Lyot plane cancellation factor of $10^{-3}$, with an occulting focal plane spot of radius $1.87 \lambda_0/D$. The Lyot stop is fixed to the diameter of the re-imaged telescope pupil. In the right-hand plot (B), the algebraic components of the Lyot plane field ($\Psi_C$) are compared, viz. Equation~\ref{eqn:Psi_C_spotfpm}. Due to the Fourier transform invariance property of the prolate-apodized field ($\Psi_A$), the filtered subtrahend component ($\Psi_A*\mathrm{jinc(\rho_0 r)}$) traces the same profile within the radius of the stop, constraining the difference ($\Psi_C$) near zero.}
\end{figure}

If the discrete apodizer solution array is $\mathcal{A}_{\rm SP}\left(r_i\right)$, where $r_i$ is the normalized pupil radius, then we can check our result by evaluating the integrated energy transmission metric originally tabulated by Soummer et al.: $2\pi\sum_{i=1}^{N_r} r_i \mathcal{A}^2_{\rm SP}\left(r_i\right)$. Our integrated energy transmission is 0.193, in close agreement with the corresponding value of the analytical prolate solution, 0.190\cite{Soummer2003circ}. A grayscale map of the apodizer transmission is plotted on the left hand side of Figure~\ref{fig:config1a}.

We decompose the two algebraic components of the Lyot plane field to learn how the design constraints are fulfilled. In the upper right plot of Figure~\ref{fig:config1a}, the apodizer curve is drawn in blue, followed by the the Hankel transform of the field inside the occulting spot, in gold. Recall that the latter curve is the function subtracted in Equation~\ref{eqn:aplc_Psi_C} to compute the resulting Lyot field. The invariance of the apodizer to the finite Hankel transform is evident by the fact that within the aperture $r < D/2$, the subtrahend curve is indistinguishable from the apodizer transmission. Outside $r = D/2$, the subtrahend remains continuous since it recovers the unrestricted prolate function. The difference of the two functions reveals a slight deviation from the analytical solution. The residual Lyot field is not shaped like the circular prolate function, as prescribed by Equation~\ref{eqn:prolate_cancellation}. Recall, however, that we did not specify a point-wise constraint in the Lyot plane, instead imposing a less stringent requirement that $ |\Psi_C(r)| < 1 - \Lambda $ for $r < D/2$.

\subsection{Focal diaphragm}

\noindent As in Equation~\ref{eqn:aplc_Psi_C} we express the on-axis Lyot field in terms of the apodizer transmission and the focal plane mask profile. To offer a slightly more intuitive description, instead of explicitly writing out the Hankel transform integrals, this time we express the Lyot plane field components in terms of pupil-plane convolutions.

\begin{equation}
\begin{aligned}
\Psi_{C_{\rm a}}(r) & = \Psi_{A_{\rm a}}(r) - \widehat{\left\{ \hat{\Psi}_{A_{\rm a}}(\rho) M_a(\rho) \right\}} \\
                              & = \Psi_{A_{\rm a}}(r) - \Psi_{A_{\rm a}}(r) * \frac{\rho_0 J_1(2\pi \rho_0 r)}{r}\\ 
\end{aligned}
\label{eqn:Psi_C_spotfpm}
\end{equation}

\begin{equation}
\begin{aligned}
\Psi_{C_{\rm b}}(\mathbf{r}) & =  \Psi_{A_{\rm b}}(r) - \widehat{\left\{ \hat{\Psi}_{A_{\rm b}}(\rho) M_b(\rho) \right\}} \\
                                             & = \Psi_{A_{\rm b}}(r) * \frac{\rho_1 J_1(2\pi \rho_1 r)}{r} -  \Psi_{A_{\rm b}}(r) * \frac{\rho_0 J_1(2\pi \rho_0 r)}{r}\\ 
\end{aligned}
\label{eqn:Psi_C_annfpm}
\end{equation}

\noindent The circular analog of the $\mathrm{sinc}$ function, $\mathrm{jinc}\left(ar\right) = a J_1\left(2\pi a r\right)/r$, appears here ($J_1$ is the first-order Bessel function of the first kind)\cite{Bracewell}. In both configurations it serves as a low-pass filter kernel on the entrance pupil field.

Equation~\ref{eqn:Psi_C_annfpm} does not yield the same form of integral equation as before, since the apodizer function appears inside an integral in both terms. Therefore, the original framework for the approximate analytical solution no longer applies. In spite of this, we show in the next subsection that a similar cancellation of the on-axis Lyot field is easily achievable with the annular FPM.

\begin{figure}[htb]
\begin{center}
\begin{tabular}{c}
\includegraphics[width=0.9\textwidth]{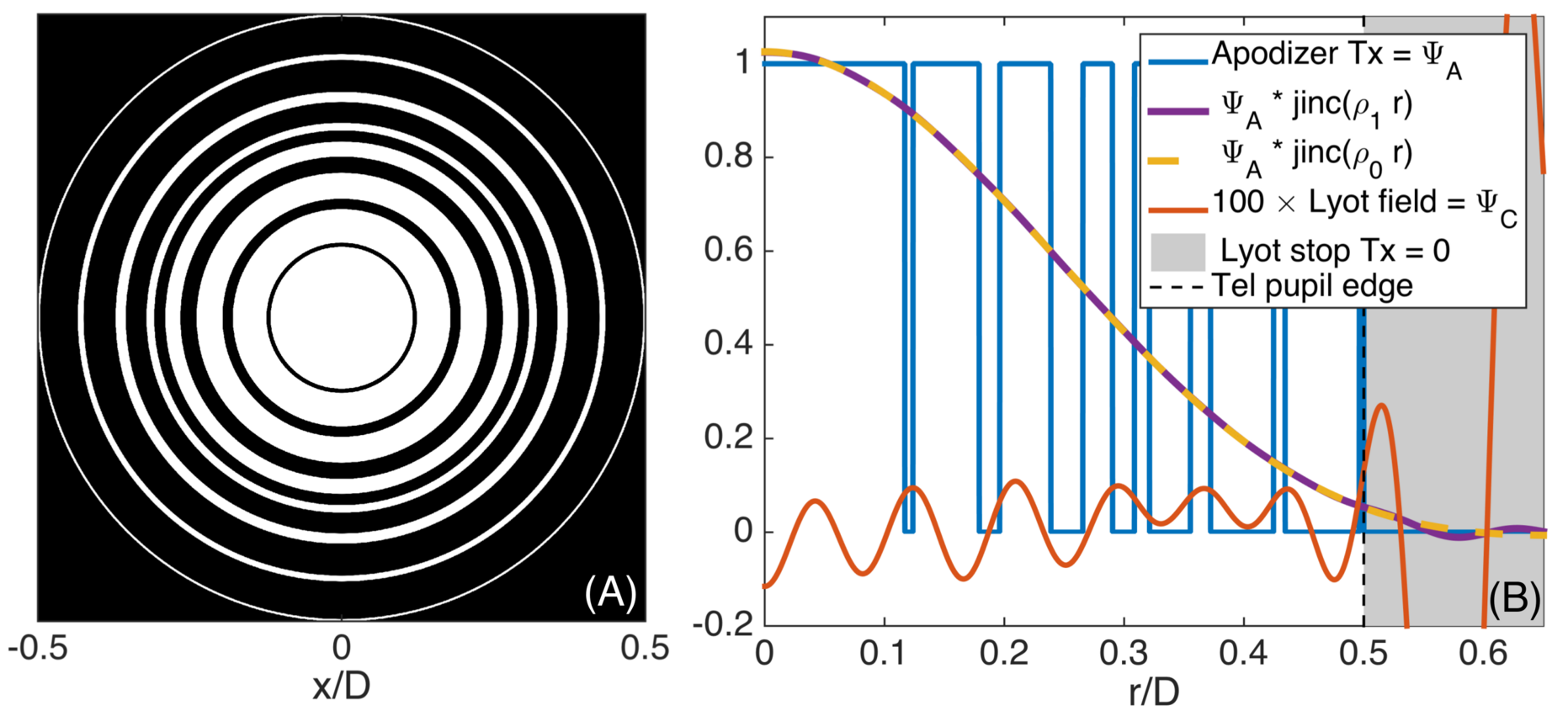}
\end{tabular}
\end{center}
\caption
{ \label{fig:config1b} 
Circular Lyot coronagraph, Config.\ Ib: The apodizer (A) is optimized for maximum transmission while achieving a monochromatic Lyot plane cancellation factor of $10^{-3}$, with an annular diaphragm FPM of inner radius $1.87~\lambda_0/D$ and outer radius $12~\lambda_0/D$. The Lyot stop is fixed to the diameter of the re-imaged telescope pupil. In the right-hand plot (B), the algebraic components of the Lyot plane field ($\Psi_C$) are compared, viz. Equation~\ref{eqn:Psi_C_annfpm}. In the image domain, $\Psi_A$ concentrates most energy within $\rho_0$ (as shown in Figure~\ref{fig:PsiB_comparison}). Therefore, both low-pass filtered instances of the ring-apodized field ($\Psi_A*\mathrm{jinc(\rho_0 r)}$ and $\Psi_A*\mathrm{jinc(\rho_1 r)}$) are approximately equal, and the residual difference meets the design constraints inside the Lyot stop.}
\end{figure}

For Config.\ Ib (Figure~\ref{fig:config1b}), we repeat the same problem as Ia, except now we replace the occulting spot with the annular diaphragm (as defined~in the second part of Equation~\ref{eqn:circ_fpm_def}). The inner and outer radii are 1.87 $\lambda_0/D$ and 12 $\lambda_0/D$, respectively. The resulting solution no longer resembles a circular prolate function, but instead a concentric ring shaped pupil mask of the kind previously described by Vanderbei et al.~\cite{Vanderbei2003}. By blocking the outer region of the focal plane, the solver is able to take advantage of a mask whose Bessel harmonics lie outside $\rho_1$, because energy distributed there can no longer propagate on to the Lyot plane. The spatial frequencies of the strong Bessel harmonics of the concentric ring mask depend on the ring spacing and thickness. As a consequence, when we repeat the trial for larger $\rho_1$, the number of rings increases, while the overall open area decreases slightly. Conversely, when $\rho_1$ is reduced, the apodizer solution has fewer, thicker rings and higher transmission.

\begin{figure}[htb]
\begin{center}
\begin{tabular}{c}
\includegraphics[height=7cm]{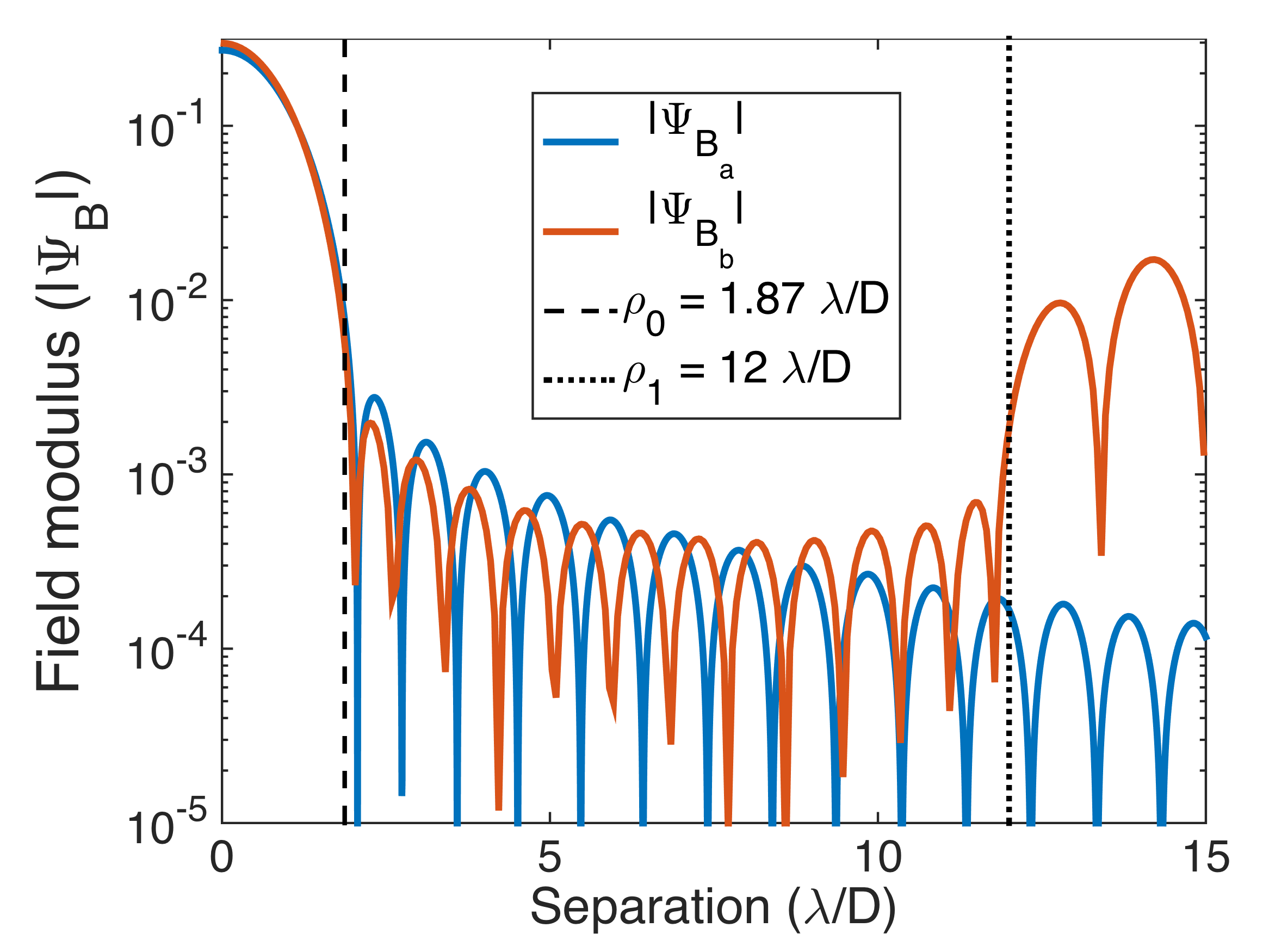}
\end{tabular}
\end{center}
\caption
{ \label{fig:PsiB_comparison}
Comparison of the scalar focal plane fields for apodizer solutions Ia and Ib. The blue curve ($\Psi_{B_a}$) is the on-axis field in the first focal plane after the circular prolate solution shown in Figure~\ref{fig:config1a}. The red curve ($\Psi_{B_b}$) is the focal plane field after the concentric ring mask apodizer of configuration Ib (Figure~\ref{fig:config1b}). Notably, the amplitude of $\Psi_{B_b}$ rises immediately outside the outer radius of the annular FPM ($\rho_1 = 12$), while the ripple envelope of $\Psi_{B_a}$ continues to fall monotonically.}
\end{figure}

The plot of the Lyot field decomposition on the right hand side of Figure~\ref{fig:config1b} reveals another interesting result. The two components of the Lyot plane field, which we expressed before in Equation~\ref{eqn:Psi_C_annfpm} in terms of convolutions between the apodizer transmission and $\mathrm{jinc}$ functions, bear a striking resemblance to the original circular prolate function that appeared in Config.\ Ia. Therefore, even though the apodizer is binary, the low-pass filter effect of the $\mathrm{jinc}$ convolution recovers a rough approximation of the circular prolate function for both the inner and outer components. The residual ripple shows the two components are equal to within the $10^{-3}$ field constraint specified in the design. This could only be the case for an apodizer that concentrates a great fraction of its energy within the inner edge of the annulus. We verify this characteristic in Figure~\ref{fig:PsiB_comparison}, where the field distributions produced in the first focal plane by the apodizers of Ia and Ib are compared. The ring apodizer has a higher overall throughput, and therefore a higher peak. Right outside the outer FPM edge, $\rho_1$, the rejected high-frequency Bessel harmonics of the ring apodizer emerge, and continue to oscillate beyond the plotted radius. By comparison, the ripple envelope of the circular prolate focal plane field decreases monotonically out to infinity.

\subsection{Polychromatic focal plane field cancellation}
\label{subsec:polychrom_FP_cancel}

While the previous trials offer a useful conceptual perspective on how binary apodizers can function in a Lyot coronagraph, we are ultimately concerned with image plane performance metrics, and solutions that suppress starlight over a finite bandwidth. Therefore, the remaining trials carry the propagation to the final focal plane and constrain the contrast there to $10^{-9}$ in a restricted region (details in Appendix~\ref{subsec:circsplc_appendix}). We define contrast here as the ratio of the intensity in the final image to the peak of the off-axis coronagraph PSF. Bandwidth is achieved by repeating the field constraints at three wavelength samples spanning a 10\% fractional bandwidth.

For all the configurations, we compute the throughput and area of the coronagraph PSF, and assemble the results in Table~\ref{tab:circ_splc_results}. Following the convention of J. Krist et al.~\cite{Krist2015JATIS}, throughput takes into account the overall proportion of energy from an off-axis (planet-like) point source that reaches the final image, as well as the proportion of that energy concentrated in the main lobe of the corresponding PSF. We make the assumption that only in the main lobe of the off-axis PSF is the intensity high enough to generate a useful signal. We compute the throughput by propagating an off-axis plane wave through the coronagraph model, masking off the full-width half-maximum (FWHM) region of the resulting PSF, and summing the intensity there. Then, we repeat the same calculation when the off-axis source is directly imaged by the telescope, without a coronagraph. The ratio of these intensity sums gives a normalized metric indicating how efficiently off-axis point sources are preserved by the coronagraph. For a Lyot coronagraph (including classical, APLC, and SPLC), throughput is constant over the field of view (FoV), as long as the off-axis PSF core clears line-of-sight obstruction by the focal plane mask.

Independent of throughput, we also assess how tightly the energy is concentrated in the central lobe of the off-axis PSF based on the area of the FWHM region. A small PSF area is desirable, because for a given throughput value, smaller area results in a higher peak signal on the detector. We again normalize this to the reference case of a PSF without a coronagraph. Since these designs have an unobstructed circular pupil, the reference telescope PSF is an Airy disk.

To enable the most meaningful comparison across the various configurations in the table, we relaxed the inner working angle from 1.87 $\lambda_0/D$ to 3 $\lambda_0/D$, where $\lambda_0$ is the center wavelength of the passband. This increase in the inner edge is needed because for some configurations, we failed to find any polychromatic solutions for $\rho_0$ of 2 $\lambda_0/D$ or below. The outer edge of the high-contrast region is arbitrarily fixed at $12~\lambda_0/D$ for all designs.

The first set of trials with image plane constraints are Configs.\ IIa and IIb, with a fixed Lyot stop again matched to the telescope aperture. Following the previous nomenclature, type ``a'' designs use the occulting spot FPM, and type ``b'' designs use the annular diaphragm FPM. For both types of focal plane mask, the apodizer with the highest throughput is a concentric ring shaped pupil (Table~\ref{tab:circ_splc_results}). Even for the spot FPM, the hard outer edge of our specified dark region means that the strong Bessel harmonics of the ring apodizer are tolerated outside $\rho_1 = 12~\lambda_0/D$. If we had constrained the contrast out to an infinite radius from the star, we would instead expect the solution to revert to a smooth apodizer with a continuous derivative. More practically, we could have included derivative constraints in the optimization program~\cite{Vanderbei2003starshaped,Kasdin2005}. In a recent APLC design study, for example, the contrast constraints were also imposed over a restricted area of the final image\cite{NDiaye2015APLCfATA4}. The authors, requiring a smooth apodizer transmission profile, added constraints on the spatial derivative of the apodizer in order to avoid binary solutions.

\begin{figure}[htb]
\begin{center}
\begin{tabular}{c}
\includegraphics[width=0.95\textwidth]{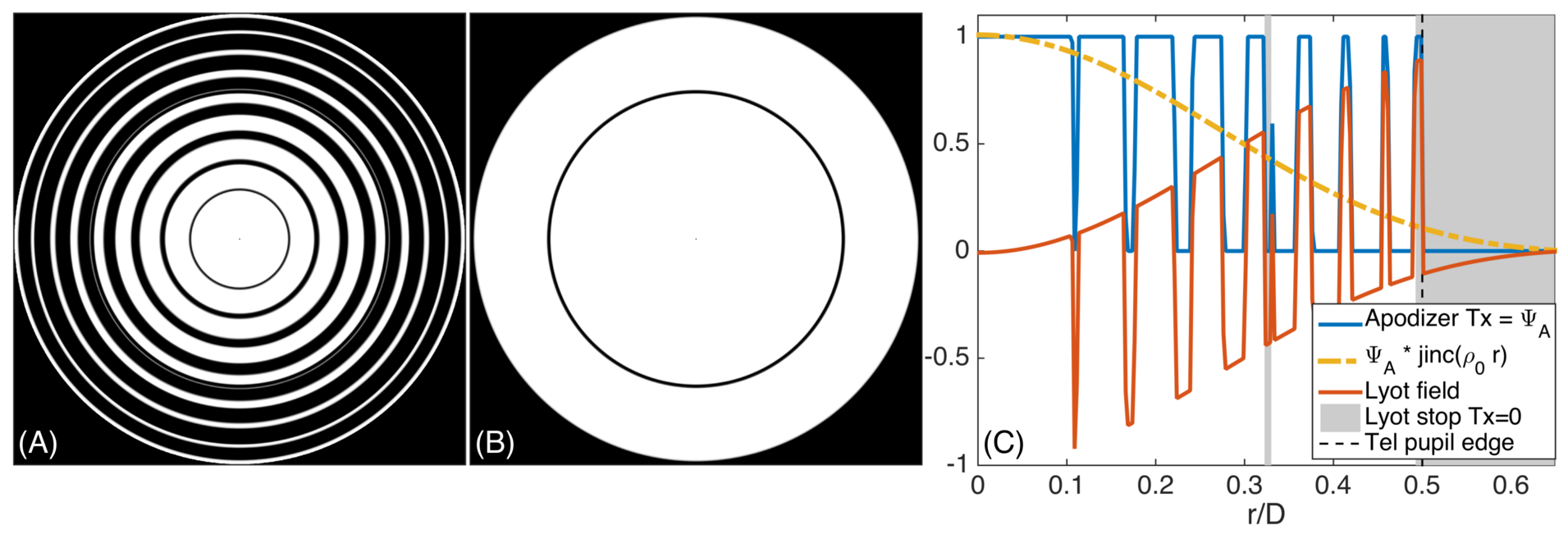}
\end{tabular}
\end{center}
\caption
{ \label{fig:config3a} Circular aperture Lyot coronagraph, Config.\ IIIa: The apodizer (A) and Lyot stop (B) are jointly optimized for maximum transmission while achieving a contrast of $10^{-9}$ in the image plane over a 10\% passband over a working angle range 2--12 $\lambda_0/D$. The focal plane mask is an occulting spot of radius $2~\lambda_0/D$. The right-hand plot (C) shows that for this configuration, the solution depends on high-amplitude discontinuities in the Lyot plane field.}
\end{figure}

\subsection{Joint optimization of the apodizer and Lyot stop}
\label{subsec:joint_opt}

For the conventional monochromatic APLC, the optimal Lyot stop is one exactly matched to the telescope aperture (after a 180 degree rotation, for telescope apertures lacking circular symmetry)\cite{Soummer2009APLCfATA2}. The Lyot stop is padded only for the purpose of alignment tolerance\cite{Soummer2011APLCfATA3}. However, recent investigations have shown that APLC optimizations incorporating bandwidth and image constraints yield better results when the Lyot stop's central obstruction replica is significantly oversized\cite{NDiaye2015APLCfATA4}. For example, in the course of optimizing an APLC for an aperture with central obstruction of diameter $0.14D$, aiming for contrast $10^{-8}$ over a 10\% bandwidth, authors N'Diaye et al.\ found that increasing the inner diameter of the Lyot stop to $\sim0.35D$ enabled the inner working angle to be reduced from $3.7~\lambda_0/D$ to $2.4~\lambda_0/D$ at the same throughput\cite{NDiaye2015APLCfATA4}. Evidently, the transmission profile of the Lyot stop offers an important parameter space to survey in addition to the apodizer.

\begin{figure}[htb]
\begin{center}
\begin{tabular}{c}
\includegraphics[height=7cm]{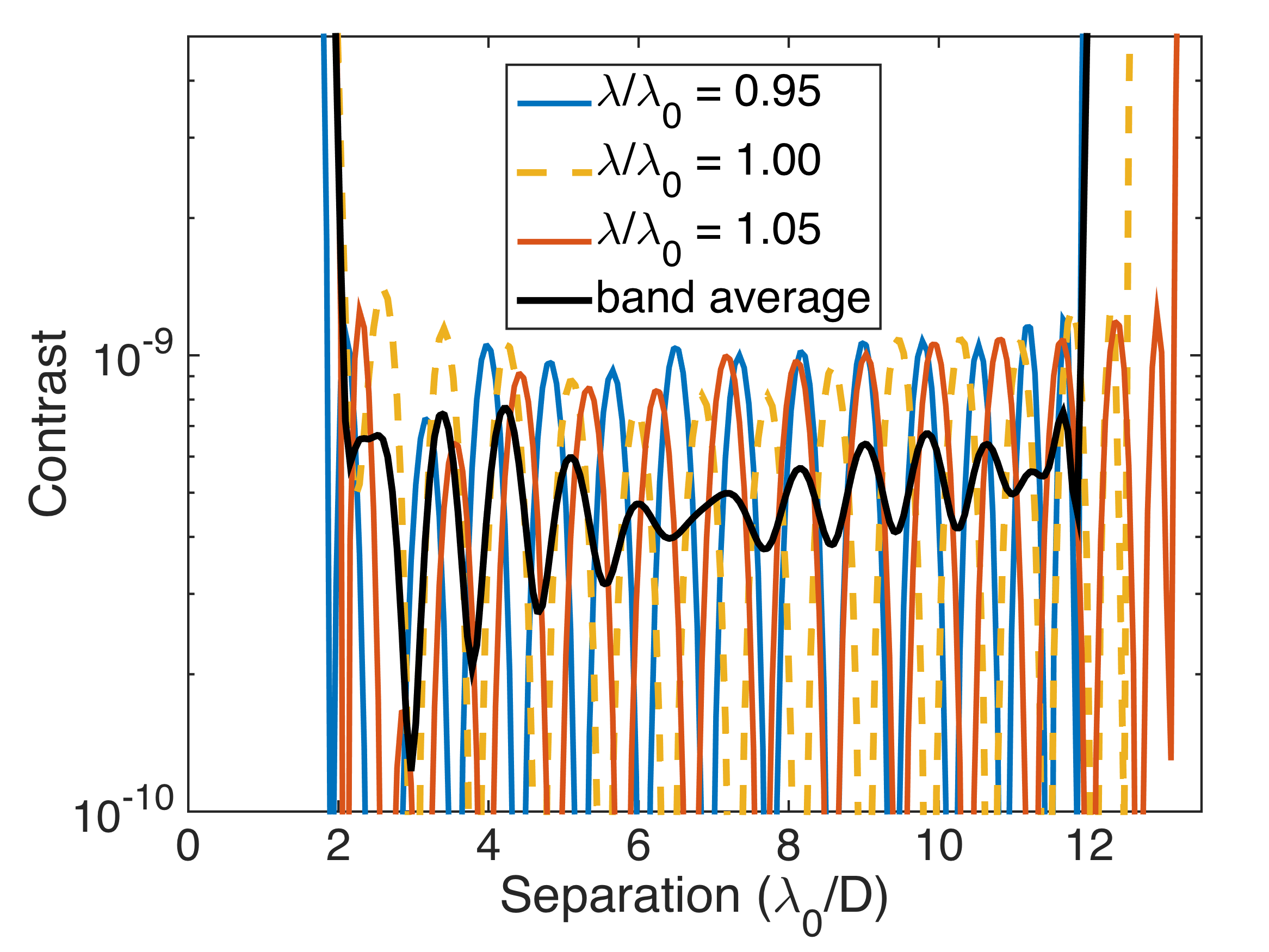}
\end{tabular}
\end{center}
\caption
{ \label{fig:polychrom_psf_config3a}
On-axis intensity pattern for a circular SPLC with a jointly optimized apodizer and Lyot stop, and working angle range 2--12$~\lambda_0/D$. The nonlinear optimization program constrains the contrast at three wavelengths, resulting in a pseudo-achromatized dark search region, fixed in sky coordinates. The black curve shows the average intensity across 5 wavelength samples spanning the 10\% passband.}
\end{figure}

\begin{figure}[htb]
\begin{center}
\begin{tabular}{c}
\includegraphics[width=0.95\textwidth]{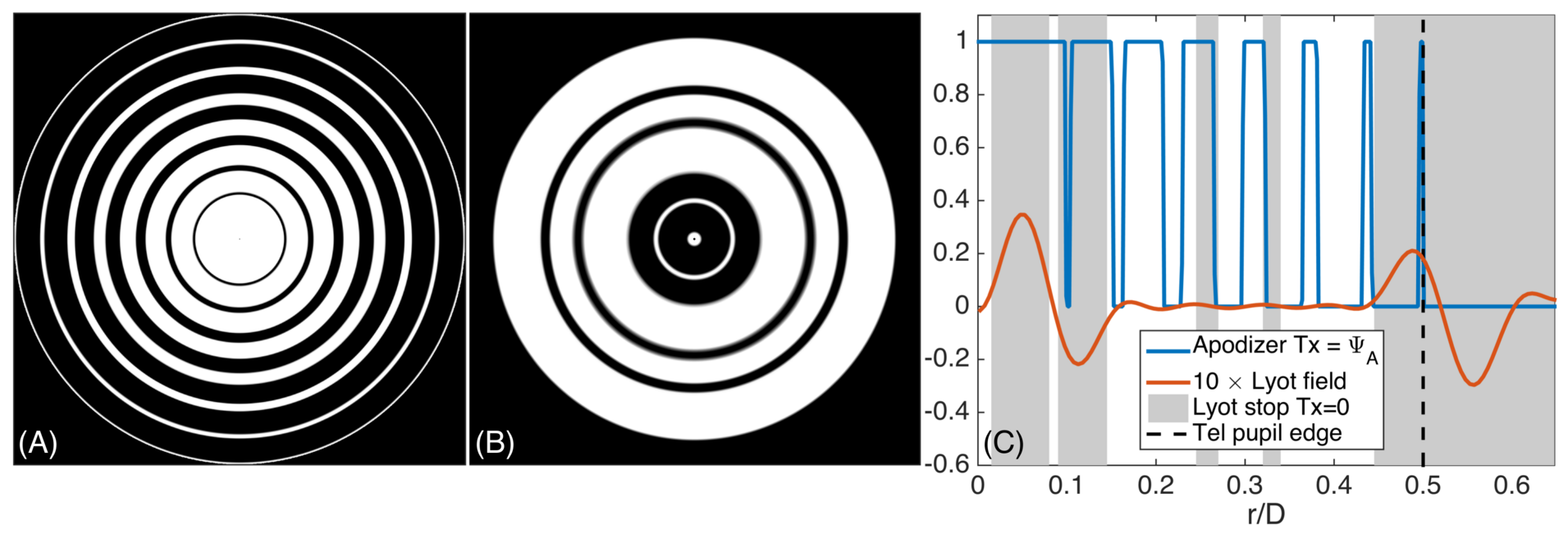}
\end{tabular}
\end{center}
\caption
{ \label{fig:config3b} Circular aperture Lyot coronagraph, Config.\ IIIb: The apodizer (A) and Lyot stop (B) are jointly optimized for maximum transmission while achieving a contrast of $10^{-9}$ in the image plane over a 10\% passband over a working angle range 2--12 $\lambda_0/D$. The focal plane mask is an annular diaphragm with inner radius $2~\lambda_0/D$ and outer radius $12~\lambda_0/D$. The Lyot plane field illustrated in the right-hand plot (C) differs from the case of Config.\ IIIa in two ways: (i) it varies smoothly, due to the filtering effect of the annular FPM; (ii) field nulls coincide with gaps in the Lyot stop.}
\end{figure}

Building on this notion, for Configs.\ IIIa and IIIb we recast the Lyot coronagraph optimization as a nonlinear program in which both the apodizer and the Lyot stop transmission profile are free vectors, optimized simultaneously. See Appendix~\ref{subsec:circsplc_appendix} for further description of this procedure. The program seeks to maximize the sum of the transmission of both apodizer and Lyot stop, given the same contrast and 10\% bandwidth goal as before. The results are illustrated in Figures~\ref{fig:config3a} and~\ref{fig:config3b}, and listed in Table~\ref{tab:circ_splc_results}. As in the case of Config.\ IIa (not plotted) the mismatch between the apodizer and the Babinet subtrahend profiles leads to high amplitude, sharp residual features in the Lyot plane. This time, however, a subtle rearrangement of Lyot stop obstructions is enough to enable an apodizer with far more open area. Most of the sharp residual Lyot plane features are not obstructed by the freely varying stop, contrary to what one might expect. Apparently, not even a modest level of field cancellation in the Lyot plane is required to create deep, broadband destructive interference in the image plane. The FWHM throughput of this solution is 0.33, more than triple that of the comparable clear Lyot stop configuration (IIa). The coronagraph PSF also sharpens, giving a FWHM area only 25\% larger than the Airy disk. We also tested the effect of decreasing the spot focal plane radius from $3~\lambda_0/D$ to $2~\lambda_0/D$, and arrived at a similar design with a throughput of 17\%. The contrast curve of this design is plotted in Figure~\ref{fig:polychrom_psf_config3a}, showing the intensity pattern at three wavelengths, as well as the average over five wavelength samples spanning the 10\% passband.

With the annular diaphragm FPM, allowing the Lyot stop transmission profile to vary also results in increased throughput, although the improvement here is less dramatic, climbing from 0.108 to 0.144 (Table~\ref{tab:circ_splc_results}). The apodizer and Lyot stop both have less open area than the spot FPM variant. Notably, however, the PSF is almost as sharp as the occulting spot variant, with FWHM area 39\% larger than the Airy disk. The plot in Figure~\ref{fig:config3b} of the Lyot plane field alongside the Lyot stop transmission profile shows that the performance of this configuration benefits from notching out the radial peaks, which was not the case for the spot FPM. The resulting Lyot stop has five prominent opaque rings and one small dark spot at the center. Another important aspect of the Lyot plane behavior for the diaphragm FPM is the relative smoothness of the field structure as compared to the spot FPM case. This is a direct outcome of the mathematical description of the Lyot field in Equation~\ref{eqn:Psi_C_annfpm}, where both instances of the apodizer transmission function are convolved with a $\mathrm{jinc}$ function. This quality of the diaphragm FPM variant of the SPLC hints at a more generous tolerance to manufacturing and alignment. We revisit this point in Section~\ref{subsec:LS_tol}, in the context of our WFIRST-AFTA designs.

From further experiments, we found that the nonlinear, nonconvex program used to derive joint shaped pupil and Lyot stop solutions only converges for one-dimensional coronagraph models. In our circular aperture case, this two-plane optimization program operates near the limit of the interior point solver's capability, and reliable outcomes require tuning. Even for low spatial resolution versions of obstructed two-dimensional apertures, there are too many variables to extend the tactic. This obstacle is algorithmic in nature rather than one that can be surmounted by expanding the computing hardware capacity. This difficulty, combined with the practical attractions of a simpler Lyot stop, suggest one might settle for an intermediate performance level by surveying an annular Lyot stop described by only two parameters (inner and outer radius). We have not yet explored the full range of inner and outer diameter Lyot stop combinations for the circular aperture. However, we found that for an arbitrary test design with a 0.1$D$ inner diameter and 0.9$D$ outer diameter (Configs.\ IVa and IVb), performance is not far from the optimized Lyot stop: for the case of the spot FPM, throughput decreases only from 0.334 to 0.317 (Table~\ref{tab:circ_splc_results}). For the diaphragm FPM variant, the throughput loss resulting from the switch to the annular Lyot stop is also small. However the coronagraph PSF deteriorates significantly, jumping in area from 1.39 to 1.93 times that of the Airy core.

\begin{table}[htb]
\caption{Summary of circular aperture Lyot coronagraph trials, for various mask and optimization configurations.}
\begin{tabular}{ c || r | r | r | c | c | l | l l }

Config.\ & Focal Plane & Lyot & Field & Band & Through & PSF & Notes \\
             & Mask & Stop & Constraint & -width & -put & area & \\ 
\hline \hline
Ia & spot & fixed & $| \Psi_C | \le 10^{-3}$ & mono & 0.134 & 1.84 & APLC for\\
  & $\rho_0 = 1.87$ & O.D. = 1 & & & & & $\Lambda = 0.999$ \\
\hline
Ib & ann $[\rho_0 = 1.87$, & fixed & $| \Psi_C | \le 10^{-3}$ & mono & 0.148 & 1.73 & 8-ring SP \\
  & $\rho_1 = 12]$ & O.D. = 1 & & & & &  \\
\hline
IIa & spot $\rho_0 = 3$ & fixed & $10^{-9}$ contrast, & 10\% & 0.095 & 2.42 & 9-ring SP \\
    & & O.D. = 1 & $3 < \rho < 12$ & & & &  \\
\hline
IIb & ann $[\rho_0 = 3$, & fixed &  $10^{-9}$ contrast, & 10\% & 0.108 & 2.18 & 8-ring SP \\
    & $\rho_1 = 12]$ & O.D. = 1 & $3 < \rho < 12$ & & & & \\
\hline
IIIa & spot $\rho_0 = 3$ & free mask & $10^{-9}$ contrast, & 10\% & 0.334 & 1.25 & 10-ring SP, \\
   &                              & & $3 < \rho < 12$ & & & & 3-ring LS \\
\hline
IIIb & ann $[\rho_0 = 3$, & free mask & $10^{-9}$ contrast, & 10\% & 0.144 & 1.39 & 7-ring SP, \\
   & $\rho_1 = 12]$       & & $3 < \rho < 12$ & & & & 7-ring LS \\
\hline
IVa & spot $\rho_0 = 3$ & fixed [I.D.=0.1, & $10^{-9}$ contrast, & 10\% & 0.317 & 1.32 & 8-ring SP \\
     &                              & O.D=0.9] & $3 < \rho < 12$ & & & & \\
\hline
IVb & ann $[\rho_0 = 3$, & fixed [I.D.=0.1, & $10^{-9}$ contrast, & 10\% & 0.121 & 1.93 & 8-ring SP \\
      & $\rho_1 = 12]$ & O.D.=0.9] & $3 < \rho < 12$ & & & & \\

\end{tabular}
\label{tab:circ_splc_results}
\end{table}

\subsection{Distinction between the SPLC and microdot realizations of the APLC}

Microdot lithography can be used to stochastically approximate the continuous prolate apodizer solutions derived from Equations~\ref{eqn:aplc_Psi_C}--\ref{eqn:prolate_cancellation}, as well as their analogs for more complicated apertures\cite{Martinez2009a,Martinez2009b,Martinez2010}. The technique stems from long-established printing processes, in which an array of black pixels with varying spatial density imitates the halftones of a grayscale image. A microdot apodizer for a Lyot coronagraph can be manufactured with an opaque metal layer deposited on a glass substrate at the locations of black pixels\cite{Martinez2009Msngr,Sivaramakrishnan2010SPIE}. In testbed experiments, APLC designs with microdot apodizers have reached contrasts as low as $5\times10^{-7}$\cite{Thomas2011AJ}. Microdot APLC apodizers are core components in several on-sky, AO-fed coronagraphs\cite{Hinkley2011PASP,Macintosh2014PNAS,Beuzit2008SPIE}.

Although the halftone microdot process results in a binary-valued transmission pattern, there is nonetheless a categorical distinction from the SPLC. A shaped pupil, rather than approximating a continuous mask solution in the apodizer plane, instead matches the desired destructive interference properties in the image plane. Consequently, on a macroscopic scale the ring apodizer shown in Figure~\ref{fig:config1b} is qualitatively dissimilar to a halftone APLC approximation, despite solving a similar field cancellation problem. Instead, the image domain is where the strongest resemblance appears between the SPLC and APLC solutions. This is made evident by comparing their on-axis field distributions at the first focal plane within the bounded search region, shown in Figure~\ref{fig:PsiB_comparison} for the most elementary design case (monochromatic cancellation in the Lyot plane).

Because the SPLC design process directly optimizes the performance, the fabrication instruction set for the apodizer realization is a one-to-one replica of the linear program solution. A microdot APLC apodizer, on the other hand, is one step removed from an underlying numerical solution. In this sense, the shaped pupil technique has a clear advantage for meeting the high precision required for the most demanding applications.
 
\section{Shaped pupil Lyot coronagraph designs for WFIRST-AFTA}

\subsection{WFIRST-AFTA CGI concept}

The Coronagraph Instrument (CGI) proposed by the WFIRST-AFTA Science Definition Team aims to image and measure the spectra of mature, long-period gas giants in the solar neighborhood. This planet population, which at present can only be studied indirectly through radial velocity surveys, is out of reach of transit spectroscopy methods due to their strong bias towards highly irradiated planets on short orbital periods. Depending on orbital configuration and albedo characteristics, the planet-to-star contrast of an exo-Jupiter seen in reflected starlight is of order $10^{-8}$ or below. Due to AO performance limitations, this contrast ratio may prove too extreme for ground-based imaging, regardless of telescope aperture or coronagraph design\cite{Guyon2005ApJ}.

The reflected spectra of gas giants are sculpted by a series of methane absorption bands in the range 600--970 nm. Acquiring these fingerprints for an ensemble of planets, in conjunction with mass constraints from radial velocities and astrometry, will provide a wealth of insights into the structure, composition, and evolution of gas giants\cite{Marley2014,Burrows2014}. The Princeton team was tasked with providing shaped pupil designs for this characterization mode, covering the stated wavelength range in three 18\% passbands, each corresponding to one filter setting of the integral field spectrograph (IFS)\cite{Noecker2015JATIS, McElwain2015JATIS, Traub2015JATIS}.

The optical path of the proposed CGI is shared between the SPC/SPLC and JPL's \textit{hybrid Lyot coronagraph} (HLC)\cite{Trauger2015JATIS}. The HLC uses a focal plane mask with a phase- and amplitude-modulating transmission profile\cite{Trauger2012SPIE}. In the baseline configuration of the CGI, the HLC mode operates with two imaging filters, nominally 10\% bandpasses centered at 465 nm and 565 nm. The HLC mode is optimized for detection and color measurements of the scattered continuum, rather than spectroscopic characterization with the wider bandpass of the IFS\cite{Traub2015JATIS}.



In addition to exoplanets, a closely related category of scientific opportunity for WFIRST-AFTA is circumstellar debris structure. One of the goals of the CGI will be to image scattered light from low-density, solar-system-like zodiacal disks that are below the noise floor of existing instruments. In addition, thick debris disks of the kind already studied with the Hubble Space Telescope will be probed at smaller angular separations than before. This will unveil the dynamical evolution of circumstellar debris and its interaction with planets in the habitable zones of exoplanetary systems\cite{Schneider2014AFTA}. Small angular separation observations of debris disks can be carried out with the HLC mode. However, some of the foreseen disk imaging programs require larger outer working angles ($>\sim$0.5 arcsec) than those relevant to reflected starlight exoplanet detection. Therefore, we explored separate SPLC mask solutions for a dedicated, wide-field ``disk science'' coronagraph mode.


Throughout our design process, we concentrate on three essential performance metrics: contrast, inner working angle (IWA), and throughput. The scientific goals require all WFIRST-AFTA designs to achieve a raw contrast of $10^{-8}$, defined at a given image position as the ratio of diffracted starlight intensity to the peak of the off-axis coronagraph PSF shifted to that location. We make the assumption that data post-processing will further reduce the intensity floor by a factor of 10 or more, so that planets several times below this nominal contrast can be detected\cite{Ygouf2015SPIE,Noecker2015JATIS, Traub2015JATIS, Krist2015JATIS}.

IWA is defined as the minimum angular separation from the star at which the coronagraph's off-axis (planet) PSF core throughput reaches half-maximum\cite{Krist2015JATIS}. For a Lyot coronagraph, planet throughput rises steadily with increasing angular separation from the edge of the FPM, leveling off when the core of the PSF clears the line-of-sight FPM occultation. Having a small inner working angle is especially important for a coronagraph aiming to detect starlight reflections, because the irradiance falls off with the square of the planet-star distance. The consequence---when considered along with the distances to nearby FGK stars and their~expected distributions of planet semi-major axes---is twofold: (i) The number of accessible planets rises steeply with reduced inner working angle; (ii) Those giant exoplanets at smaller angular separations tend to be the brightest targets\cite{Traub2014SPIE,Traub2015JATIS}.

As in Section~\ref{sec:circ_splc} we define throughput as the ratio of energy contained within the FWHM contour of the PSF core, to that of the telescope PSF with no coronagraph. Planet signal-to-noise ratio will generally be low, and the number of targets the instrument can acquire over the mission lifespan will be limited by the cumulative integration times\cite{Traub2015JATIS}. Detection times will depend on the total amount of planet light that survives propagation losses through the optical train, and how tightly that remaining energy is concentrated on the detector. In characterization mode, the spectrograph will disperse the planet's light over many detector pixels. Therefore, the majority of the instrument's operational time budget will be consumed by integrations totaling one day or more per target\cite{Krist2015JATIS}.

\begin{figure}[htb]
\begin{center}
\begin{tabular}{c}
\includegraphics[height=5cm]{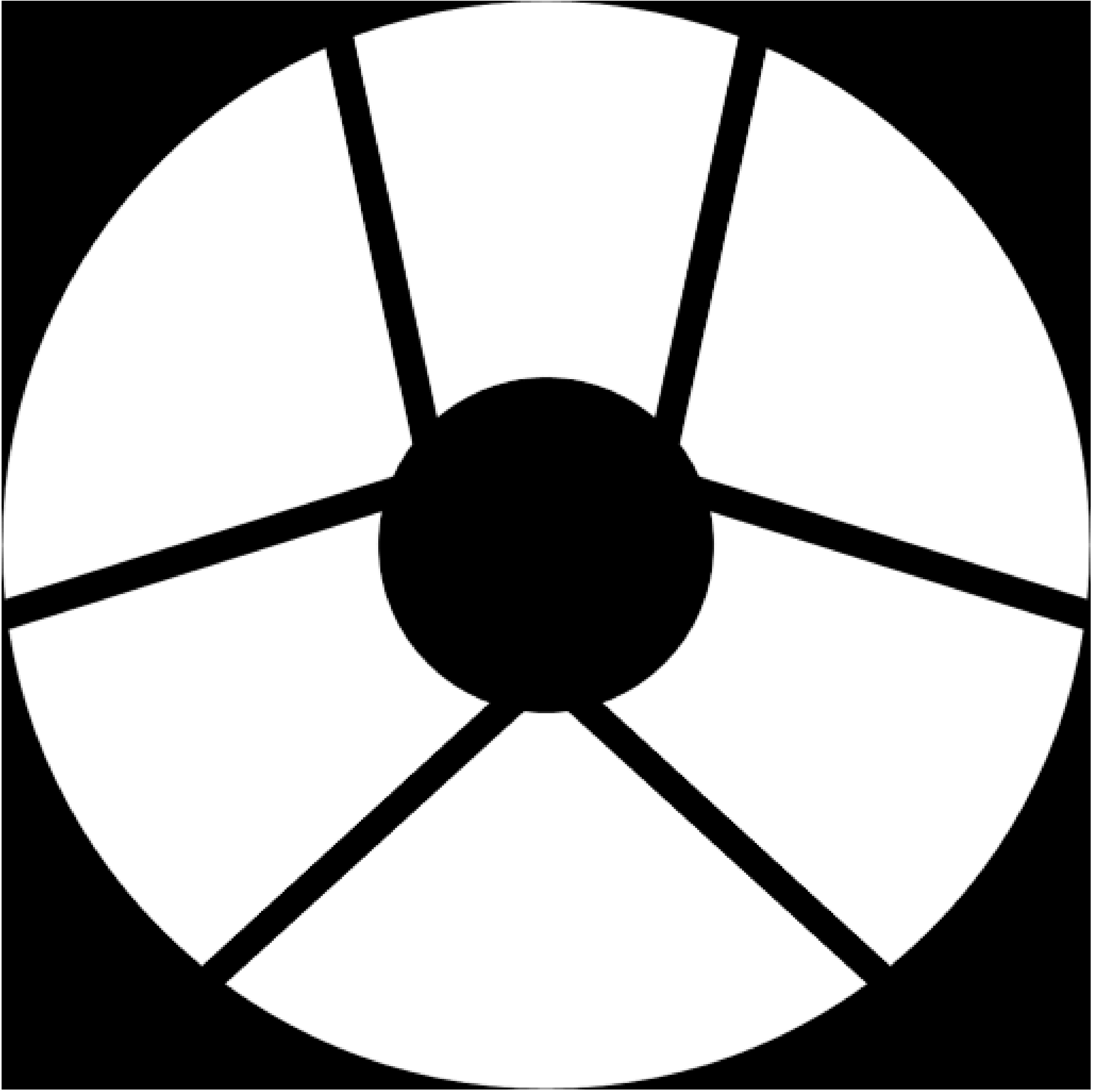}
\end{tabular}
\end{center}
\caption
{ \label{fig:afta_pupil}
The WFIRST-AFTA telescope aperture from the most recent mission design cycle, as seen on-axis.}
\end{figure}

The 2.4 m-diameter WFIRST-AFTA telescope aperture is illustrated in Figure~\ref{fig:afta_pupil}. Its large central obstruction (0.31$D$) and six off-center support struts, each oriented at a unique angle, pose a challenge for any coronagraph design. The SPC and SPLC use the apodization pattern of the shaped pupil to confine the diffraction effects of these obstructions outside of the optimized dark region\cite{Carlotti2012, Carlotti2013AFTA}. The penalties of this strategy are lower throughput and higher inner working angle than would be the case for a clear circular aperture. Alternatively, it is possible for a coronagraph to counteract these obstructions with static phase excursions applied to deformable mirrors\cite{Pueyo2013ACAD, Trauger2015JATIS} or custom aspheric optics\cite{Guyon2005PIAA,Abe2006,Guyon2012SPIE,Pluzhnik2015JATIS}. However, amplitude-mask-based apodization places less demanding requirements on mirror surface manufacturing tolerances, alignment tolerances, and deformable mirror reliability, thereby mitigating the overall technological risk of our design.

The SPLC designs presented here build directly on the efforts of A. Carlotti et al.~\cite{Carlotti2013AFTA}, who led the first shaped pupil designs for WFIRST-AFTA. Those ``first generation'' shaped pupil coronagraphs fulfill the basic mission requirements. They were described in further detail by A. J. E. Riggs et al.~\cite{Riggs2014SP}, and in this issue E. Cady et al.\ describe successful laboratory demonstrations of the first-generation characterization SPC design~\cite{Cady2015JATIS}. The SPLC designs here form part of a reference design case adopted by the Science Definition Team for the purpose of technology demonstrations, mission simulations, and cost assessment\cite{Spergel2015WFIRST}. The flight design moving forward may differ significantly.

\subsection{Characterization mode SPLC}
\label{subsec:afta_char_splc}

For the characterization design, the challenge is to achieve a small inner working angle while maintaining acceptable throughput. Although it is always desirable to create a full 360-degree dark search region around the star, we know from previous work that it is impossible for a shaped pupil alone to produce an annular FoV with IWA 4 $\lambda/D$ or below with the obscurations of the WFIRST-AFTA aperture\cite{Carlotti2013AFTA}. However, knowing that the SPLC configuration should be able to reach a smaller IWA at the same contrast and throughput as the first generation SPC, we now examine again how close we can push a 360-degree dark region in towards the star. At the same time, for effective broadband characterization we strongly prefer a quasi-achromatic dark region\cite{Soummer2011APLCfATA3}, so that a target located near the IWA is detected across the full spectrograph passband. Therefore, for our parameter exploration we always apply polychromatic image constraints, with inner and outer image radii defined in terms of central wavelength diffraction elements ($\lambda_0/D$), as we did before in Section~\ref{subsec:polychrom_FP_cancel}, and similar to previous APLC optimizations described by N'Diaye et al.\cite{NDiaye2015APLCfATA4}. In Appendix~\ref{subsec:afta_appendix} we describe the practical details of the optimization procedure used to test a given set of design parameters.

Informed by the results of our circular SPLC trials (Sections~\ref{subsec:polychrom_FP_cancel}--\ref{subsec:joint_opt}), we use an occulting spot focal plane mask, with radius either $\rho_0 = $2.5 or 3.0 $\lambda_0/D$. We survey the Lyot stop parameter space by repeating optimizations with different padding levels on the inner and outer edge of the telescope aperture replica, ranging from to 2\% to 12\% of the diameter. We also varied the outer radius of the dark region between 8, 9, and 10 $\lambda_0/D$. Finally, we repeated these multi-parameter trials for two bandwidths: 18\% (the target characterization design) and 10\%. A subset of the results are summarized in Table~\ref{tab:char360}. Several conclusions can be drawn. First, for the smaller inner FPM radius $\rho_0 = 2.5~\lambda_0/D$, there are no acceptable 360-degree SPLC solutions for the full 18\% characterization bandwidth. At $\rho_0 = 3.0~\lambda/D$, however, some weak solutions begin to appear. Performance here is sensitive to the Lyot stop padding~level, and for the 18\% bandwidth case the best padding levels are in the range 8--10\% of the pupil diameter. When the bandwidth is reduced to 10\%, the improvement in throughput is dramatic. In particular, we highlight a design with a throughput of 0.14 and FPM radius $3.0~\lambda_0/D$. We did not find a strong dependence on outer dark region radius $\rho_1$ over the values we surveyed.

\begin{table}[htb]
\center
\caption{Throughput of 360-degree field-of-view WFIRST-AFTA SPLC solutions for different combinations of inner working angle, bandwidth, and Lyot stop padding. The outer edge of the dark image constraint region is fixed at $\rho_1 = 8~\lambda_0/D$.}
\begin{tabular}{ c || c | c || c | c }
 & \multicolumn{2}{ c ||}{18\% bandwidth} & \multicolumn{2}{ c }{10\% bandwidth} \\
 & \multicolumn{2}{ c ||}{$\rho_0$ ($\lambda_0/D$)} & \multicolumn{2}{ c }{$\rho_0$ ($\lambda_0/D$)} \\
 \hline
 \hline
Lyot stop padding (\% diam.) & 2.5 & 3.0 & 2.5 & 3.0 \\
\hline
6 & $<10^{-3}$ & 0.042 & 0.040 & 0.13  \\
8 & $<10^{-3}$ & 0.057  & 0.060 & \textbf{0.14} \\
10 & $<10^{-3}$ & 0.066 & 0.062 & 0.12 \\
12 & $<10^{-3}$ & 0.056 & 0.043 & 0.07 \\
\end{tabular}
\label{tab:char360}
\end{table}

To reach an inner working angle smaller than 3 $\lambda_0/D$, and do so over the full characterization bandwidth, we need to restrict the azimuthal span of the constrained dark region. This strategy was originally developed to design the first-generation WFIRST-AFTA SPCs, resulting in a design with $\rho_0 = 4\lambda/D$. In a~survey aiming to discover exoplanets, bowtie-shaped dark zones have the disadvantage of requiring repeat integrations for two or more mask orientations. However, the overhead for characterizing a planet with a known position is minor. Furthermore, restricting the azimuthal and radial FOV of the image plane search area can give better suppression with wavefront correction. When the wavefront control system aims to suppress light only in a small region, there are more degrees of freedom available than when trying to suppress the full correctable region. Restricting the dark hole problem thus leaves greater tolerance for unknown aberrations in the propagation model.

\begin{figure}[htb]
\begin{center}
\begin{tabular}{c}
\includegraphics[width=0.9\textwidth]{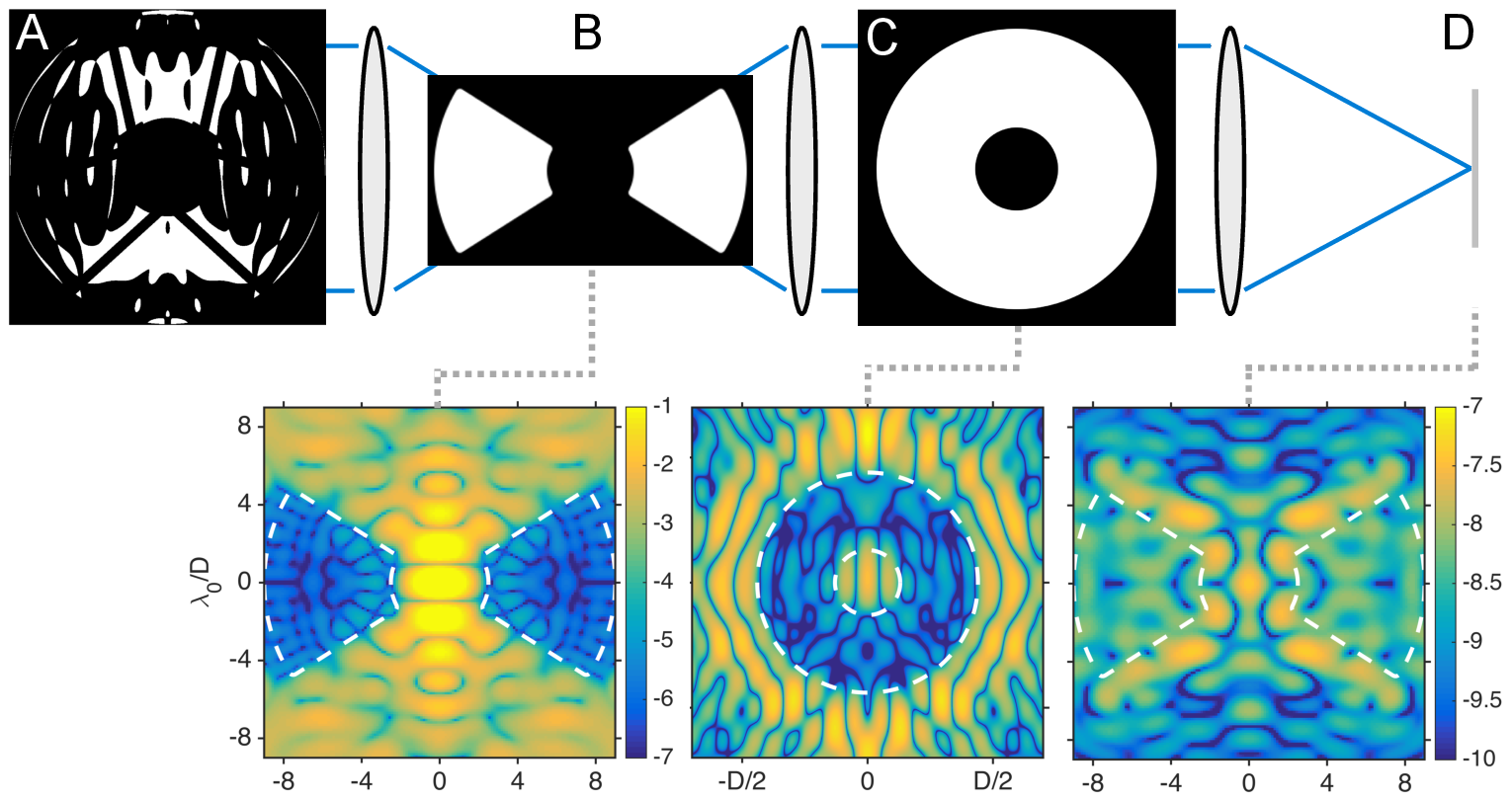}
\end{tabular}
\end{center}
\caption
{ \label{fig:char_mask_schem}
Diagram of the characterization-mode SPLC mask scheme, along with the plots of the intensity of the on-axis field at each critical plane. The shaped pupil (A) forms a bowtie-shaped region of the destructive interference in the first focal plane (B), which is then occulted by a diaphragm with a matched opening. The on-axis field is further rejected by an annular stop in the subsequent Lyot plane (C), before it is re-imaged at the entrance of the integral field spectrograph (D). The propagation is shown at the central wavelength of the design, for the case of a perfectly flat wavefront with no planet or disk present. The flux scale bars indicate the intensity on a log$_{10}$ scale. In the first focal plane (B), the flux scale is normalized to the PSF peak, whereas in the final focal plane (D) the scale is normalized to the peak of the unseen off-axis PSF, in order to map the contrast ratio in a way that accounts for the Lyot stop attenuation. The mean contrast in the dark bowtie region (averaged over azimuth angle, then averaged over wavelength, and then averaged over radial separation) is $6\times10^{-9}$.}
\end{figure}

For the SPLC configuration, we surveyed a range of bowtie-shaped focal plane geometries with inner radii between $2.4~\lambda_0/D$ and $3.0~\lambda_0/D$, and opening angles between 30 and 90 degrees. The trials are repeated for 18\% and 10\% bandwidths. All optimization attempts at smaller inner radii, such as $2.2~\lambda_0/D$, failed to give results with reasonable throughput (above $0.01$). Here, instead of an occulting spot, the focal plane mask is a bowtie-shaped aperture matched to the optimized focal plane region in the final image, as illustrated in Figure~\ref{fig:char_mask_schem}. The presence of the outer edge in the first focal plane makes this design most analogous to Config.\ IVb, among the circular SPLCs described in Section~\ref{sec:circ_splc} and Table~\ref{tab:circ_splc_results}.

The Lyot stop we use for the bowtie characterization design is a simple clear annulus rather than a padded replica of the telescope aperture (Figure~\ref{fig:char_mask_schem}). That is because the low-pass filter effect of the bowtie FPM smears the support strut field features in the Lyot plane. We verified through separate optimization tests that there is no advantage to be had by including matched support struts in the Lyot stop in this configuration. To survey the dependence of throughput on focal plane geometry, we fix the inner diameter of the Lyot stop annulus at $0.3D$ and the outer diameter at $0.9D$. Later, we tune the inner and outer diameters for a specific characterization focal plane shape.

Some results from the focal plane geometry trials are collected in Table~\ref{tab:char_FP_survey}. Since we found that throughput depends relatively weakly on outer dark region radius, here we only tabulate the throughput values for $\rho_1  = 9~\lambda_0/D$. Throughput varies steeply with opening angle. In particular, from 60 deg to 90 deg, the throughput decreases by a factor of $\sim$4--5. At inner radius $2.4~\lambda_0/D$, the only opening angle~with throughput above 0.1 is the 30-degree bowtie.

We find the most compelling trade-off at $\rho_0 = 2.6~\lambda_0/D$, which at opening angle 60 degrees has throughput 0.11. Similar to the first generation SPC, this enables the full FOV to be covered with three pairs of shaped pupils and FPMs oriented at 120-degree offsets. Due to their limited utility, the 10\% bandwidth trials are not tabulated here, but we can summarize them by pointing out that throughput increases by a factor of 1.1 to 3 over the 18\% bandwidth case, with the largest changes occurring for small $\rho_0$ and wide opening angle.

\begin{table}[htb]
\center
\caption{Throughput of bowtie characterization WFIRST-AFTA SPLC solutions for different combinations of inner radius and opening angle. The optimization passband is 18\%, the outer radius is fixed at $\rho_1 = 9~\lambda/D$, and the Lyot stop is a clear annulus with [I.D., O.D.] = [0.3, 0.9]. The approximate tradeoff in inner radius, opening angle, and throughput we chose to pursue is highlighted in bold font.}
\begin{tabular}{ c || c | c | c | c | c }
 & \multicolumn{5}{ c }{ Opening angle (deg) } \\
 \hline
 $\rho_0$ ($\lambda_0/D$) & 30 & 45 & 60 & 75 & 90 \\
 \hline
2.4 &  0.14 &   0.089 &  0.048 & 0.026 & 0.010 \\
2.6 &  0.16 &   0.17 &  \textbf{0.11} &  0.049 & 0.022 \\
2.8 &  0.16 &   0.18 &  0.13 &  0.067 & 0.030 \\
3.0 &  0.17 &   0.18 &  0.15 &  0.088 & 0.042 \\
\end{tabular}
\label{tab:char_FP_survey}
\end{table}

\begin{figure}[htb]
\begin{center}
\begin{tabular}{c}
\includegraphics[height=7cm]{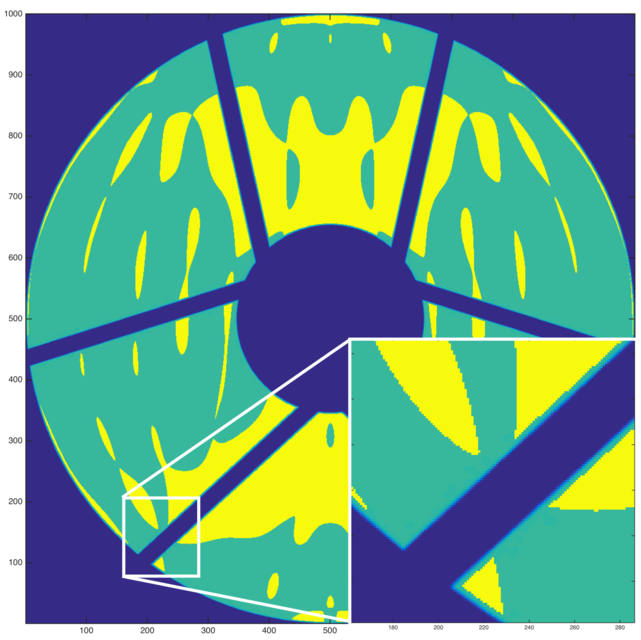}
\end{tabular}
\end{center}
\caption
{ \label{fig:char_SP_mask_detail}
Detail of the $1000\times1000$ point shaped pupil mask solution for the WFIRST-AFTA characterization mode, corresponding to the design exhibited in Figures~\ref{fig:char_mask_schem} and~\ref{fig:char_psf_and_contrast}. The obscurations of the WFIRST-AFTA telescope aperture are colored blue; the regions of the pupil masked by the shaped pupil apodizer in addition to the telescope pupil are colored green; and the regions of the pupil transmitted by the apodizer are colored yellow. The magnified inset shows the granular quality of the square, binary elements of the shaped pupil array. The inset also shows the gap between the edge of the telescope aperture features and the open regions of the apodizer, which is reserved in order to ease the alignment tolerance between the shaped pupil apodizer and the telescope pupil.}
\end{figure}

A full set of mask designs for a bowtie characterization SPLC have been delivered to JPL for fabrication and experiments on the High Contrast Imaging Testbed. The detailed structure of the $1000\times1000$ pixel shaped pupil apodizer array is illustrated in Figure~\ref{fig:char_SP_mask_detail}.

For this version, we raised the opening angle above 60 deg to provide a margin of FoV overlap between the three mask orientations needed to cover an annulus around the star. The overlap slightly reduces the likelihood of a scenario where the location of an exoplanet coincides with the edge of a bowtie mask, cropping the PSF core and requiring extra integration time to compensate. After the opening angle was fixed at 65 deg, we decremented $\rho_0$ from 2.6~$\lambda_0/D$ to the smallest radius that maintains the throughput above an arbitrary goal of 0.10, thereby revising $\rho_0$ to 2.5~$\lambda_0/D$. The Lyot stop is an annulus with inner diameter $0.26D$ and outer diameter $0.88D$. We stress that these design choices are provisional and that maximizing the scientific yield would require integrating the parameter survey with end-to-end observatory and data simulations.

\begin{figure}[htb]
\begin{center}
\begin{tabular}{c}
\includegraphics[width=0.95\textwidth]{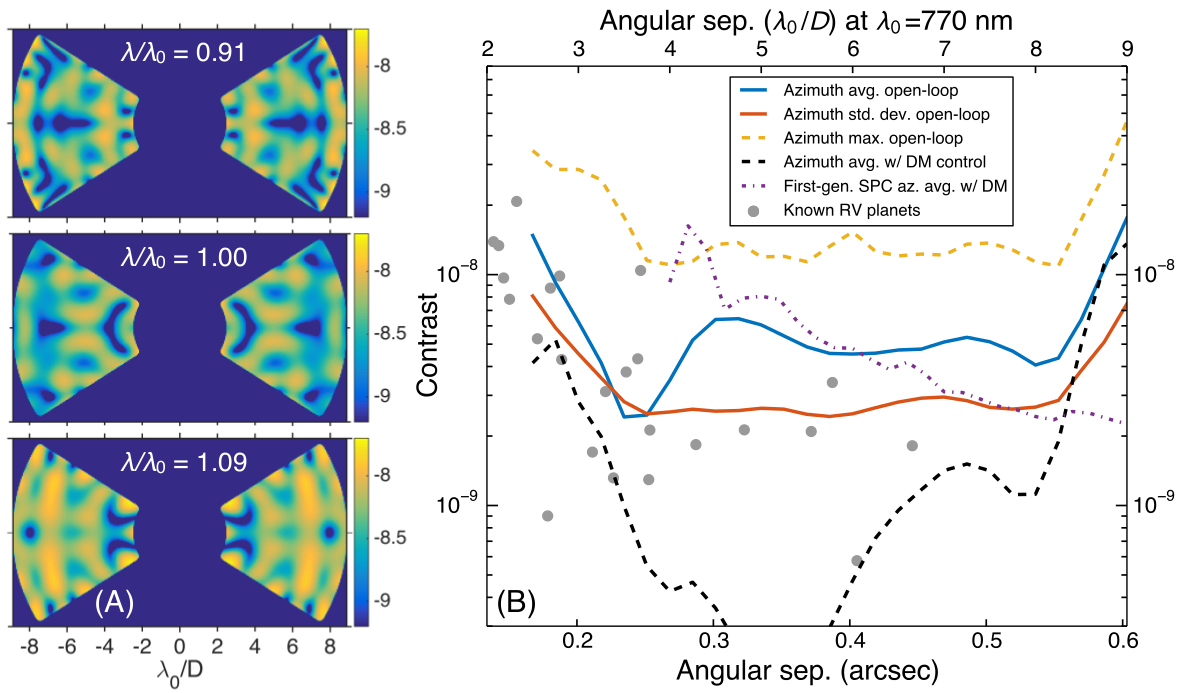}
\end{tabular}
\end{center}
\caption
{ \label{fig:char_psf_and_contrast}
(A) Ideal on-axis PSF of the characterization SPLC design, with no aberrations, shown at three wavelengths: the short wavelength extreme of the passband (top; $\lambda/\lambda_0 = 0.91$), the center wavelength (middle; $\lambda/\lambda_0 = 1.00$), and the long wavelength extreme (bottom; $\lambda/\lambda_0 = 1.09$). (B) Raw contrast plotted alongside the estimated contrasts and separations of the known long-period radial velocity exoplanet population. The contrast curves are shown for the SPLC design centered on $\lambda_0 = 770$~nm, the middle of the three spectrograph filters. For reference, we also include the contrast curve of the first-generation SPC (the purple dashed curve).}
\end{figure}

The ideal model PSF and contrast curves with zero wavefront error are shown in Figure~\ref{fig:char_psf_and_contrast}. At center wavelength, the FWHM PSF area is 1.6 times that of the WFIRST-AFTA PSF. The contrast constraint is slightly relaxed relative to the previous parameter trials: $2\times10^{-8}$ for separations below 3.5 $\lambda_0/D$, and $1.5\times10^{-8}$ in the rest of the bowtie region. Still, the average contrast curve (here averaged over azimuth and then over wavelength) is well below the worst-case intensity values, as plotted in the right hand side of Figure~\ref{fig:char_psf_and_contrast}, below $7\times10^{-9}$ at all angular separations in the range 3--8 $\lambda_0/D$.

The two deformable mirrors integrated with the WFIRST-AFTA CGI are expected to improve the nominal SPLC performance. Since the SPLC design optimization only makes use of amplitude operations, the extra degrees of freedom from DM phase control can yield higher contrast. To demonstrate this, we simulate the effect of DM control with an unaberrated wavefront. We simulated wavefront control on a layout similar to the actual WFIRST-AFTA CGI with two $48{\times}48$-actuator DMs upstream of the SPLC. We divided the 18\% passband into nine wavelength samples and weighted each equally to control the dark hole with a stroke minimization algorithm, originally described by Pueyo et al.~\cite{pueyo2009optimal} The inner region of the bowtie is most critical since more exoplanets are expected to be observed at small angular separations, so we weighted the intensity from $2.5-4.5~\lambda_0/D$ three times higher for a slight improvement. The resulting contrast curve is plotted in the right hand side plot of Figure~\ref{fig:char_psf_and_contrast}. With DM control, the average intensity in the separation range 2.5--3.5 $\lambda_0/D$ is reduced by a factor of 2, and at separations 4--8 $\lambda_0/D$ by a factor of 4 or more. In addition to the azimuthally averaged contrast, in the same figure we also plot the standard deviation of the intensity pattern as a function of separation from the star, measured in concentric annuli.

We use the distribution of known radial velocity exoplanets to indicate where the performance of this coronagraph sits relative to plausible characterization targets. We take the same assumptions made by W. Traub et al.\ in Ref.~\citenum{Traub2015JATIS} in their science yield calculations, resulting in a representative target population plotted on the right hand side of Figure~\ref{fig:char_psf_and_contrast}. Planets are assumed to be on circular orbits inclined by 60 degrees, and observed at a favorable mean anomaly of 70 degrees. The longest period RV exoplanet detections are generally on the upper end of the mass distribution, so in the absence of other constraints they are assigned a size equal to Jupiter, and a geometric albedo of 0.4. Finally, to compare the SPLC contrast curve on an angular scale, we set the central wavelength of the characterization design to 770 nm, the middle of the three nominal IFS filters\cite{McElwain2015JATIS}. It can be seen that 12 planets outside of the $2.8~\lambda_0/D$ inner working angle have contrasts and separations placing them above the band-averaged contrast floor obtained from the wavefront control simulation.

The first-generation characterization SPC design for WFIRST-AFTA is overplotted as the dashed purple contrast curve in Figure~\ref{fig:char_psf_and_contrast}. With an IWA of $4\lambda/D$, that coronagraph could only access half of the exoplanets in the mock target sample. There is an additional disadvantage of the first-generation design that is not apparent in the contrast plot. The inner working angle of a shaped pupil PSF scales directly with wavelength. Without a Lyot stop, there is no possibility to anchor the inner radius of the dark bowtie region across the spectrograph filter bandpass, as we do to optimize the SPLC. Therefore, an exoplanet falling near the inner edge of the first-generation contrast curve would be undetected at the long-wavelength end of the filter.

 We acknowledge that the raw contrast prediction for the characterization SPLC is optimistic since they do not include aberrations, tip-tilt jitter, or alignment errors, etc. But the comparison verifies that our design functions in the regime of contrast and angular separation needed to meet the top-level mission requirement of acquiring the reflected spectra of 6 or more gas giants\cite{Traub2015JATIS}. In this issue, J. Krist et al.\ analyze the sensitivity of the SPLC performance to realistic aberrations, and incorporate the coronagraph in an end-to-end simulation of the WFIRST-AFTA telescope for a complete observing scenario\cite{Krist2015JATIS}.

\subsection{Debris disk mode SPLC}
\label{subsec:afta_disk_splc}

\begin{figure}[htb]
\begin{center}
\begin{tabular}{c}
\includegraphics[width=0.9\textwidth]{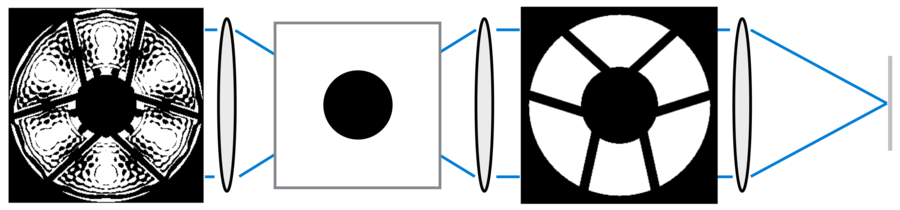}
\end{tabular}
\end{center}
\caption{\label{fig:disk_mask_schem} Mask scheme for the debris disk mode SPLC for WFIRST-AFTA, from left to right: shaped pupil apodizer, focal plane mask, and Lyot stop. The relative sizes of the masks in conjugate planes are not shown to scale. This design produces a broadband, annular dark region with~$\le10^{-8}$ contrast over an 18\% bandwidth from $6.5~\lambda_0/D$ to $20~\lambda_0/D$. The focal plane mask is an occulting spot of radius $6.5~\lambda_0/D$. The Lyot stop is a replica of the telescope aperture, with the inner and outer edges padded at 4\% of the pupil diameter.}
\end{figure}

To design an SPLC for debris disk imaging over a much wider FoV, we carry out a parameter survey similar to that of the 360-degree characterization design trials. Assuming a deformable mirror with a $48\times48$ actuator array, the maximum correctable aberration spatial frequency corresponds to angular separation $21.8~\lambda_0/D$ at the short wavelength end of an 18\% passband. To approximately match this, we fix the outer radius of the polychromatic dark region at $20~\lambda_0/D$. Within that dark annulus we constrain the contrast to $\le10^{-8}$ over an 18\% bandwidth. We test FPM spot radii of 6.0, 6.5, and 7.0 $\lambda_0/D$, and Lyot stop padding levels between 2\% and 8\%. The throughput results are tabulated in Table~\ref{tab:disk_params}. We highlight the solution with  $\rho_0 = 6.5~\lambda_0/D$ and padding level 4\%, since it gives a throughput of 0.23, almost as high as the best design at $\rho_0 = 7~\lambda_0/D$. With the smaller focal plane radius of $\rho_0 = 6~\lambda_0/D$, on the other hand, there is a significant throughput drop for all the Lyot stops.

\begin{table}[htb]
\center
\caption{Throughput of debris disk science WFIRST-AFTA solutions for different combinations of inner radius and Lyot stop padding. The tradeoff in inner radius and throughput corresponding to the design exhibited in Figures~\ref{fig:disk_mask_schem} and~\ref{fig:disk_psf_and_contrast} is highlighted in bold font.}
\begin{tabular}{ c || c | c | c }

 & \multicolumn{3}{ c }{$\rho_0$ ($\lambda_0/D$)} \\
 \hline
 \hline
Lyot stop padding (\% diam.) & 6.0 & 6.5 & 7.0 \\
\hline
2 & 0.17 & 0.21 & 0.22 \\
4 & 0.18 & \textbf{0.23} & 0.24 \\
6 & 0.17 & 0.20 & 0.22 \\
8 & 0.14 & 0.15 & 0.17 \\
\end{tabular}
\label{tab:disk_params}
\end{table}

\begin{figure}[htb]
\begin{center}
\begin{tabular}{c}
\includegraphics[width=0.95\textwidth]{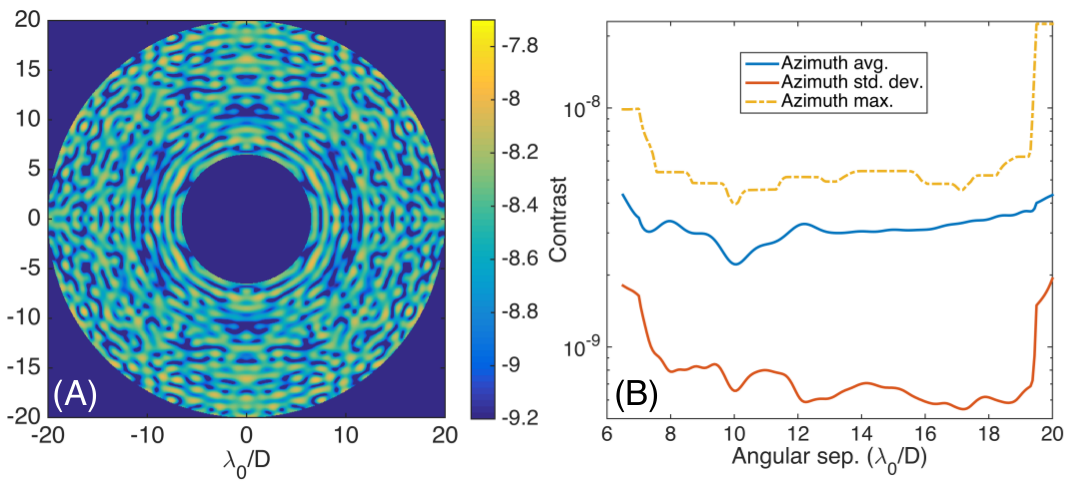}
\end{tabular}
\end{center}
\caption
{ \label{fig:disk_psf_and_contrast}
A) The ideal, on-axis, center wavelength PSF of a broadband disk science SPLC design. On the right hand side (B) the ideal contrast averaged over azimuth samples and then averaged over 7 wavelength samples spanning the 18\% passband. Also plotted are the maximum contrast at each separation, over all azimuth and wavelength samples.}
\end{figure}

In Figure~\ref{fig:disk_mask_schem} we illustrate the SPLC mask scheme for the highlighted disk science design. The apodizer maintains over 59\% of the available open area around the WFIRST-AFTA pupil obscurations, and the FWHM PSF area is only 1.11 times that of the WFIRST-AFTA telescope. The on-axis PSF of the coronagraph at the center wavelength is plotted on the left hand side of Figure~\ref{fig:disk_psf_and_contrast}, along with the ideal contrast curves. Like the characterization design, the mean contrast is significantly deeper than the constraint value, due to the lumpy structure of the diffraction pattern.

We have also optimized debris disk designs that use an annular diaphragm FPM, instead of an occulting spot. The throughput is in this configuration is approximately half that of the spot FPM case for the same contrast, FoV, and bandwidth parameters. However, a general advantage of SPLC designs that use the annular diaphragm FPM is their greater tolerance to Lyot stop mask misalignment, an issue we discuss in Section~\ref{sec:tolerances}. Therefore, despite the lower theoretical performance, for the initial testbed implementation of the debris disk SPLC we will use an annular FPM variant\cite{Zimmerman2015SPIE}.

\subsection{Summary of WFIRST-AFTA SPLC designs}

We assemble in Table~\ref{tab:afta_splc_summary} the parameters and performance metrics of our candidate SPLC designs for WFIRST-AFTA. To be consistent with other coronagraph descriptions, the inner and outer working angles (IWA and OWA) are measured by the half-maximum crossings of the throughput curve, rather than the dimensions of the FPM and optimization constraints\cite{Guyon2006ApJS}. We use the same definitions for throughput and PSF area first given in Section~\ref{subsec:polychrom_FP_cancel}. The PSF area is the FWHM region of the PSF for an off-axis (planet-like) source, normalized to the FWHM area of the PSF of the WFIRST-AFTA telescope without a coronagraph. We note that two designs are listed for the debris disk mode. One is the occulting spot configuration described in Section~\ref{subsec:afta_disk_splc}, Figures~\ref{fig:disk_mask_schem} and~\ref{fig:disk_psf_and_contrast}; the other is the annular diaphragm variant that is undergoing fabrication for testbed evaluation at HCIT, described in more detail in Ref.~\citenum{Zimmerman2015SPIE}.

\begin{table}[htb]
\caption{Summary of WFIRST-AFTA SPLC designs.}
\begin{tabular}{ c || c | c | c | c | c | c | c }

Configuration & IWA & OWA & FoV & Contrast & Band & Through & PSF \\
            & ($\lambda_0/D$) & ($\lambda_0/D$) &       &         & -width        & -put        & area \\ 
\hline \hline
Characterization mode & 2.8 & 8.7 & 2$\times$65~deg & $6.0\times10^{-9}$ & 18~\% & 0.10 & 1.6 \\
\hline
Disk mode / spot FPM & 6.6 & 19.9 & 360 deg & $3.2\times10^{-9}$ & 18~\% & 0.22 & 1.1 \\
\hline
Disk mode / ann. FPM & 6.8 & 19.7 & 360 deg & $3.1\times10^{-9}$ & 10~\% & 0.15 & 1.2 \\ 
\end{tabular}
\label{tab:afta_splc_summary}
\end{table}

\section{Error tolerances}
\label{sec:tolerances}

\subsection{Shaped pupil apodizer}

For the first-generation SPC designs for WFIRST-AFTA, Riggs et al.\ found that the contrast performance was sufficiently robust to etching errors\cite{Riggs2014SP}. The current testbeds in the JPL HCIT are using shaped pupil masks with diameters between 14 mm and 22 mm. With 1000 optimized transmission points across the mask diameter, the binary array is therefore composed of square ``pixels'' of width 14--22 microns. The standard etching tolerance in JPL's Microdevices Lab (MDL) is below 1 micron, so the etching error on each SP pixel is less than about 5\%. The first-generation characterization SPC only had a nominal, open-loop contrast degradation of 2$\times$, from 1 to 2$\times10^{-8}$, for a 5\% uniform over/under-etching error. For the mission payload, even tighter etching tolerances can be achieved, so uniform etching errors are not a major concern for the SP apodizer mask.

\subsection{Focal plane mask and pointing sensitivity}

For the bowtie characterization SPLC described in Section~\ref{subsec:afta_char_splc}, we modeled the sensitivity to systematic errors in the focal plane mask fabrication and alignment, as well as line-of-sight pointing errors originating from spacecraft jitter. The width of the stellar PSF core in relation to the inner radius of the bowtie mask is highest at the long wavelength end of the design passband. Any disparity from the transmission profile assumed in the on-axis optimization therefore causes unwanted starlight to leak into the inner part of the dark region. Conversely, at the short wavelength end of the passband, the outer perimeter of the dark bowtie region is sensitive to transmission profile disparities along the outer edge of the bowtie mask.

\begin{figure}[htb]
\begin{center}
\begin{tabular}{c}
\includegraphics[height=8cm]{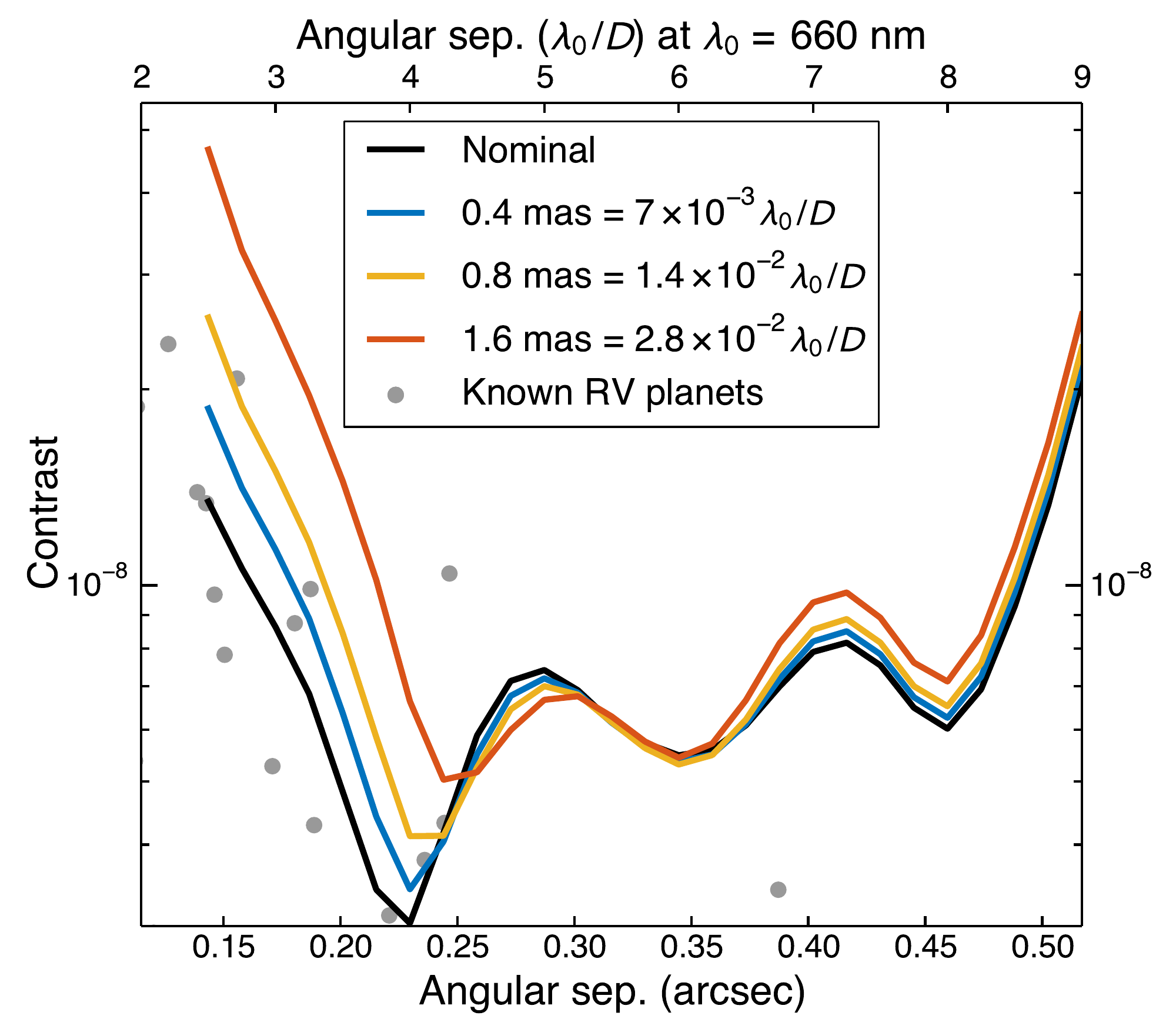}
\end{tabular}
\end{center}
\caption
{ \label{fig:pointing_error_contrast}
Contrast degradation of WFIRST-AFTA characterization design for three pointing errors, shown at $\lambda = 719$ nm, the red end of the design centered on the $\lambda_0 = 660$ nm spectrograph filter.}
\end{figure}

We plot the effect of pointing errors, and in particular their effect on the inner part of the image, in Figure~\ref{fig:pointing_error_contrast}. In our Fourier propagation model, we apply phase ramps in the apodizer plane corresponding to a set of tilt errors along the long axis of the bowtie, for the characterization design with central wavelength 660 nm. This wavelength corresponds to the bluest of the three characterization filters, chosen here because when IWA is fixed in resolution elements ($\lambda_0/D$), the shortest wavelength coronagraph realization---as defined by the physical scale of the FPM---is the one most sensitive to a given telescope pointing error. We apply pointing errors of 0.4, 0.8, and 1.6 milliarcsec. These are on the same angular scale as the residual jitter levels that may be present on the WFIRST spacecraft\cite{Noecker2015JATIS, Traub2015JATIS, Krist2015JATIS}. At the 660 nm center wavelength of the characterization passband, these tilts translate to focal plane offsets of $7\times10^{-3}~\lambda_0/D$, $1.4\times10^{-2}~\lambda_0/D$, and $2.8\times10^{-2}~\lambda_0/D$, respectively. To show the upper bound of the impact on the polychromatic coronagraph PSF, in Figure~\ref{fig:pointing_error_contrast} we plot the contrast only at the long-wavelength end of the passband, 719 nm, where the effect is worst. At each separation, we plot the azimuthal average of the contrast in the degraded half of the bowtie region.

The results plotted in Figure~\ref{fig:pointing_error_contrast} show that contrast outside of $4~\lambda_0/D$ is not degraded significantly for pointing errors up to 1.6 milliarsec, roughly equivalent to an inner FPM radius error of $\sim3\times10^{-2}~\lambda_0/D$. At the interior, the long-wavelength intensity increments for 0.4, 0.8, and 1.6 milliarcsec are respectively $6\times10^{-9}$, $1.4\times10^{-8}$, and $3.6\times10^{-8}$, as expressed in units of contrast. We emphasize that these values only indicate the contrast degradation at the red edge of the characterization filter. The impact over the rest of the band is much smaller: with a 1.6 milliarcsec pointing error, for example, the intensity increment at the interior of the bowtie averaged over 7 wavelength samples is $3.6\times10^{-9}$, a factor of 10 below the increment at the red extreme. It remains to be seen through integrated modeling on the full scope of the observatory, including low-order wavefront sensing and control, and data post-processing, how tip-tilt error affects the science yield of the characterization SPLC. For example, there may be an advantage in re-designing the apodizer for a passband extended slightly beyond the actual spectrograph filter in order to provide a buffer against pointing or alignment errors, at some cost in throughput.

Errors in the lateral alignment between the apodizer and the FPM have a comparable impact on the on-axis intensity as line-of-sight pointing errors. We summarize the results of trial offsets in Table~\ref{tab:char_FPM_trans_tol}. We quantify this in two ways. First, by computing the mean change in contrast over all the spatial samples and wavelength samples in the dark bowtie region. Then, we isolate the region of the image that is most severely affected: the inner part of the bowtie (within $3.8~\lambda_0/D$), at the red edge of the bandpass. Offsets up to $4\times10^{-2}~\lambda_0/D$ cause a mean contrast degradation of only 2--4$\times10^{-9}$. However, in the most sensitive part of the image, for a horizontal offset of $2\times10^{-2}~\lambda_0/D$ the contrast increment approaches $10^{-8}$.

\begin{table}[htb]
\center
\caption{Contrast degradation of the characterization SPLC for a set of horizontal and vertical translation offsets of the focal plane mask.}
\begin{tabular}{ l || c | c }
& mean $\Delta$Contrast; & mean $\Delta$Contrast; \\
& all spatial and  & sep. $< 3.8~\lambda_0/D$, \\
& wavelength samples & red edge of 18\% band,  \\
&  & worst quadrant \\
\hline
horz. $5\times10^{-3}~\lambda_0/D$ & $3.0\times10^{-11}$ & $1.5\times10^{-9}$ \\
horz. $1\times10^{-2}~\lambda_0/D$ & $1.2\times10^{-10}$ & $3.4\times10^{-9}$ \\
horz. $2\times10^{-2}~\lambda_0/D$ & $4.8\times10^{-10}$ & $8.5\times10^{-9}$ \\
horz. $4\times10^{-2}~\lambda_0/D$ & $1.9\times10^{-9}$ & $2.4\times10^{-8}$ \\
\hline
vert. $5\times10^{-3}~\lambda_0/D$ & $5.8\times10^{-11}$ & $3.2\times10^{-9}$ \\
vert. $1\times10^{-2}~\lambda_0/D$ & $2.3\times10^{-10}$ & $8.0\times10^{-10}$ \\
vert. $2\times10^{-2}~\lambda_0/D$ & $9.3\times10^{-10}$ & $2.3\times10^{-9}$ \\
vert. $4\times10^{-2}~\lambda_0/D$ & $3.8\times10^{-9}$ & $7.6\times10^{-9}$ \\
\end{tabular}
\label{tab:char_FPM_trans_tol}
\end{table}

We have also tested the effect of bowtie FPM clocking errors on the characterization SPLC performance. The clocking angle of the bowtie FPM in the first focal plane needs to be accurate to within 0.5 degrees to keep the worst contrast degradation (over all spatial samples and wavelengths) below $5\times10^{-9}$.

\subsection{Lyot stop}
\label{subsec:LS_tol}

We test the Lyot stop alignment tolerance of the WFIRST-AFTA characterization SPLC by modeling the propagation of flat, on-axis wavefronts when the Lyot stop is translated off-center. Over 7 wavelength samples and 27 angular separations spaced at $\left(\lambda_0/D\right)/4$, we compute the azimuthally averaged contrast. Then we compute the mean and maximum increment relative to the nominal contrast values, over all those wavelengths and separations. We find that the coronagraph performance is more sensitive to horizontal than vertical Lyot stop translations (here horizontal and vertical orientations are used in the same sense as the diagram in Figure~\ref{fig:char_mask_schem}). In Table~\ref{tab:charsplc_LS_tol} we summarize the effect for horizontal translations 0.5\%, 1.0\%, and 2.0\% of the pupil diameter.

\begin{table}[htb]
\center
\caption{Contrast degradation of WFIRST-AFTA Characterization SPLC for horizontal Lyot stop alignment errors.}
\begin{tabular}{ c || c | c }
Lyot stop translation (\% pupil diam.) & mean $\Delta$Contrast & max. $\Delta$Contrast \\
\hline
0.5 & $2.1\times10^{-10}$ & $1.3\times10^{-9}$  \\
1.0 & $9.2\times10^{-10}$ & $5.5\times10^{-9}$ \\
2.0 & $4.7\times10^{-9}$ & $2.5\times10^{-8}$  \\
\end{tabular}
\label{tab:charsplc_LS_tol}
\end{table}

The focal plane mask of the WFIRST-AFTA characterization SPLC is opaque outside the optimized bowtie region (Figure~\ref{fig:char_mask_schem}). Any such diaphragm in the first focal plane has a low-pass filter effect on the morphology of the field in the Lyot plane, as we showed in Equation~\ref{eqn:Psi_C_annfpm} for the simple circular case. The effect can also be examined visually in the evaluation plots of the earlier ``type b'' SPLC configurations in Section~\ref{sec:circ_splc}. The lack of sharp field transitions near the Lyot stop edges helps to keep the alignment tolerance reasonable, in spite of the very high contrast goal in the final image.

This smooth Lyot field characteristic is not the case for the WFIRST-AFTA debris disk design we presented in Section~\ref{subsec:afta_disk_splc}, which operates with an occulting spot FPM. Initial calculations indicate that its Lyot stop alignment tolerance is at least an order of magnitude tighter than that of the diaphragm FPM variants. It is possible that an expanded optimization procedure can counteract this sensitivity. For example, if the optimizer model propagates the field not only through perfectly aligned masks, but also through a set of cases with translated Lyot stops, then the final field could be constrained simultaneously for misaligned mask scenarios. However, if that approach is not feasible, then there is a substantial practical advantage for SPLC designs optimized for a diaphragm-type FPM, despite the fact that their throughput is in most cases lower for the same image constraints (e.g., comparing the metrics of Configs.\ IVa and IVb in Table~\ref{tab:circ_splc_results}, and the debris disk designs in Table~\ref{tab:afta_splc_summary}).

\section{Conclusion}

We have described a hybrid coronagraph configuration that uses a shaped pupil as the apodizing mask in a Lyot-style architecture. An optimized SPLC reaches the contrast and inner working angle of the well-established APLC design family, while benefitting from a precise, achromatic transmission characteristic that is most feasible with a binary apodizer\cite{Bala2015JATIS}.

Our numerical optimization experiments have revealed a rich parameter space in the Lyot stop transmission profile. The apodizer and Lyot stop can be optimized simultaneously, leading to solutions with higher throughput and a sharper PSF for a given contrast and bandwidth. For example, we noted one design (Figure~\ref{fig:polychrom_psf_config3a}) that surpasses $10^{-9}$ contrast starting from angular separation $2~\lambda_0/D$, while maintaining a full-width half-max throughput of 17\% over a 10\% passband. At present, however, due to optimizer limitations the approach is only feasible for telescope apertures with pure circular symmetry.

The SPLC is compatible with two types of focal plane mask: a conventional occulting spot and an annular diaphragm. Once the Lyot stop is tuned, the throughput is generally higher for the occulting spot solutions. However, alignment and manufacturing tolerances may hinder their practicality, due to sharp field features in the Lyot plane originating from the binary profile of the apodizer. By distinction, the low-pass filter effect of the diaphragm FPM dramatically relaxes the tolerance on the Lyot stop profile accuracy. Future efforts will determine if an expanded optimization procedure can produce occulting spot solutions that are less sensitive to this effect.

By applying the same design principles tested for the circular case, we explored the parameter space of SPLC solutions for WFIRST-AFTA. We arrived at a mask scheme optimized for the atmospheric spectroscopy mode of the coronagraph. This design produces a bowtie-shaped ($2\times65$ deg), quasi-achromatic dark region of $<10^{-8}$ contrast over an 18\% bandwidth, with an inner working angle of $2.8~\lambda_0/D$ (0.19 arcsec at $\lambda_0 = 770$ nm). Experiments at JPL HCIT are underway to test the ability of this coronagraph to meet the exoplanet characterization goals of the mission\cite{Cady2015JATIS}. We are also evaluating promising designs for a wider-angle disk imaging mode, operating from $6.5~\lambda_0/D$ to $20~\lambda_0/D$ (angular separations 0.4--1.5 arcsec over the IFS filter set).

We limited the practical aspects of this study to the WFIRST-AFTA mission concept, but SPLC designs have broad applicability to high-contrast imaging problems with obscured telescope apertures. Upcoming work by M. N'Diaye et al. will survey SPLC solutions for different aperture geometries and scientific goals.

\appendix
\section{Optimization scheme}

\subsection{Circular aperture SPLC}
\label{subsec:circsplc_appendix}

For each circular SPLC configuration, a discrete, algebraic propagation model enables us to exactly define the optimization objectives and constraints we explored in Section~\ref{sec:circ_splc}. We mimic the notation used in past descriptions of conventional (non-Lyot) shaped pupil coronagraph optimizations\cite{Carlotti2011,Vanderbei2012}. As in those cases, we code the algebraic model and design goals as a linear program in the AMPL programming language. For each of the circular SPLC experiments, we used the LOQO interior point solver\cite{VanderbeiLOQO} to solve the AMPL program and obtain the mask solution.

Due to the circular symmetry of the telescope pupil, the on-axis (stellar) scalar field in each coronagraph plane is expressed as a purely real, one-dimensional, radial function. The numerical Fourier propagation between the coronagraph planes is computed via the discrete Hankel transform. We use spatial coordinate $r_i$ in the re-imaged telescope pupil, and $\xi_j$ in the image plane. The image coordinate maps to a true physical radius, and does not scale with wavelength. A unitless wavelength ratio, $\gamma_k = \lambda_k/\lambda_0$, where $\lambda_0$ is the center wavelength, captures the chromatic dependence of the field.

We use the variable $\mathcal{A}_{\rm SP}$ to represent the radial transmission function of the shaped pupil apodizer and $\mathcal{A}_{\rm LS}$ to represent the Lyot stop. The variables $\Psi_B$, $\Psi_C$, and $\Psi_D$ represent the scalar fields in the first focal plane, Lyot plane, and final focal plane, respectively. For pupil plane variables (namely, $\mathcal{A}_{\rm SP}$, $\mathcal{A}_{\rm LS}$, and $\Psi_C$), the radial coordinate $r_i$ is normalized to the re-imaged telescope pupil diameter. Therefore, if there are $N_R$ points across the pupil radius, spaced at interval $\Delta r = \frac{1}{2}/N_R$, then the radial samples occur at $r_i = \left(i - \frac{1}{2}\right)\Delta r$, for integers $i=1,2,\dots,N_R$.

\subsubsection{Focal occulting spot}

In the case where the FPM is an occulting spot (Configs.\ Ia, IIa, IIIa, and IVa in Section~\ref{sec:circ_splc}), we use the semi-analytical APLC modeling approach of Soummer et al.\cite{Soummer2007}. In the first focal plane, we compute the field only within the occulting spot, rather than the (ideally) unbounded transmitted region. Then Babinet's superposition principle can be applied to determine the Lyot plane field, as expressed before in Equation~\ref{eqn:aplc_Psi_C}. We sample the field at $N_{\rho_0}$ points within the spot radius $\rho_0$, at spacing $\Delta\xi = \rho_0/N_{\rho_0}$. Then for integers $j = 1,2,\dots,N_{\rho_0}$, we compute the interior field at image radii $\xi_j = (j - \frac{1}{2}) \rho_0/N_{\rho_0}$, in units of center wavelength resolution elements. The expressions for the field in the first focal plane and the Lyot plane follow:

\begin{equation}
\begin{aligned}
\Psi_{B}\left(\xi_j, \gamma_k\right) =& 2\pi/\gamma_k \sum\limits_{i=1}^{N_R} r_{i} \mathcal{A}_{\rm SP} \left(r_i\right) J_0\left(2\pi \xi_j r_i /\gamma_k \right) \Delta r, \\
\Psi_C \left(r_i, \gamma_k\right) =& \mathcal{A}_{\rm SP}\left(r_i\right) - 2\pi/\gamma_k \sum\limits_{j=1}^{N_{\rho_0}} \xi_j J_0 \left(2\pi r_i \xi_j / \gamma_k\right) \Psi_{B}\left(\xi_j, \gamma_k\right) \Delta \xi. \\ 
\end{aligned}
\end{equation}

\noindent In our Section~\ref{sec:circ_splc} trials, the sampling interval is as fine as $\Delta\xi=\frac{1}{16}$ in the first focal plane, and $\Delta r=\frac{1}{2}/2000$ in the pupil planes. For Config.\ Ia, the Lyot plane $\Psi_C$ is the last stage of propagation computed by the optimizer. As in the case of the conventional APLC, the on-axis field is constrained here\cite{Soummer2003circ}. Our goal is to maximize the sum of the apodizer mask field transmission over the pupil area, while meeting some level of on-axis field cancellation. Since the design is monochromatic, $\Psi_C$ is only computed and constrained at $\gamma_k=1$. Now we have the elements needed to declare the optimization objective, along with the design constraints:

\begin{equation}
\label{eqn:obj_config1a}
\begin{aligned}
\mathrm{Maximize}~\mathcal{O}_I &= 2\pi \displaystyle\sum\limits_{i=1}^{N_R} r_i \mathcal{A}_{\rm SP}\left(r_i\right) \Delta r,\\
\mathrm{subject~to:}~0 &\le \mathcal{A}_{\rm SP}\left(r_i\right) \le 1,~\mathrm{and}\\
-10^{-s}~&\le~ \Psi_C\left(r_i\right) \le ~10^{-s},~\mathrm{for}~0 \le r_i \le \frac{1}{2}.\\
\end{aligned}
\end{equation}

\noindent The parameter in the field cancellation exponent, $s$, is set to 3.0 in the case illustrated in Section~\ref{sec:config1a}.

In order to more directly prescribe the performance, as we do for Configs.\ IIa--IVa, the optimization model must propagate the field from the Lyot plane to the final focal plane of the coronagraph. Here, we switch the spatial coordinate variable from $\xi$ to $\zeta$ to indicate a change in the radial sampling. The new sampling interval, $\Delta\zeta$, must be no larger than $\frac{1}{2}$ of a center wavelength resolution element (to meet the Nyquist-Shannon sampling criterion) and preferably close to $\frac{1}{4}$. The expression for the scalar electric field in the final plane is

\begin{equation}
\Psi_D\left(\zeta_j, \gamma_k \right) = 2\pi/\gamma_k \sum\limits_{i=1}^{N_R} r_i \mathcal{A}_{\rm{LS}}\left(r_i\right) \Psi_C\left(r_i, \gamma_k\right) J_0\left(2\pi \zeta_j r_i /\gamma_k\right) \Delta r, \\
\end{equation}

\noindent computed at radii $\zeta_j = \left(j - \frac{1}{2}\right)\Delta\zeta$ for integer indices $j$ satisfying $\rho_0 \le \zeta_j \le \rho_1$. For Config.\ IIa/IIb, the Lyot stop is a replica of the telescope pupil, therefore $\mathcal{A_{\rm LS}}$ is equal to unity for all radii in the summation bounds. However, for Config.\ IVa/IVb, the annular Lyot stop is equal to unity for $0.1 \le r_i/\frac{1}{2} \le 0.9$ and zero elsewhere.

Our goal again is to maximize the integrated field transmission of the apodizer mask $\mathcal{A}_{\rm SP}$. This time, however, the on-axis field is constrained in an annular region of the final image. For each wavelength ratio $\gamma_k$ sampling the operating bandwidth, we compute the peak field in the first focal plane. This value is a proxy for the star's peak intensity, and serves as a reference for the contrast constraints:

\begin{equation}
\Psi_{B_{\rm Peak}}\left(\gamma_k\right) = 2\pi/\gamma_k \displaystyle\sum\limits_{i=1}^{N_R} r_i \mathcal{A}_{\rm SP}\left(r_i\right) \Delta r. \\
\end{equation}

\noindent Now we can declare the optimization objective and the design constraints:

\begin{equation}
\label{eqn:obj_config2a}
\begin{aligned}
\mathrm{Maximize}~\mathcal{O}_I &= 2\pi \displaystyle\sum\limits_{i=1}^{N_R} r_i \mathcal{A}_{\rm SP}\left(r_i\right) \Delta_r,\\
\mathrm{subject~to:}~0 &\le \mathcal{A}_{\rm SP}\left(r_i\right) \le 1,~&\mathrm{for}~0 \le r_i \le \frac{1}{2},~\mathrm{and}\\
-10^{-c/2}~&\le~ \frac{\Psi_D\left(\zeta_j, \gamma_k\right)}{\Psi_{B_{\rm Peak}}\left(\gamma_k\right)} \le ~10^{-c/2},~&\mathrm{for}~\rho_0 \le \zeta_j \le \rho_1\mathrm{~and~}\\
&~ &1-w/2 \le \gamma_k \le 1+w/2.\\
\end{aligned}
\end{equation}

\noindent The $c$ parameter in the exponent is the base 10 logarithm of the desired contrast in intensity. In practice, $c$ must be increased slightly above this specification to compensate for the off-axis field attenuation caused by the Lyot stop. The parameter $w$ defining the wavelength bounds is the fractional operating bandwidth (equal to 0.1 for most trials in Section~\ref{sec:circ_splc}). By repeating identical constraints at multiple wavelength samples, the true spatial dimensions of the dark search region (and equivalently, its angular projection on the sky) are fixed across the operating bandwidth. Similar achromatization procedures have been applied to APLC designs\cite{Soummer2011APLCfATA3,NDiaye2015APLCfATA4}. At the 10\% bandwidth we investigated for the circular aperture, three wavelength samples suffice to maintain a broadband null at $10^{-9}$ contrast.

If, as in configuration IIIa, we define the Lyot stop as a variable rather than a fixed parameter, then the optimization objective must take into account the transmission of two masks. In our trials described in Section~\ref{subsec:joint_opt}, we weight them equally:

\begin{equation}
\label{eqn:obj_config3a}
\begin{aligned}
\mathrm{Maximize}~\mathcal{O}_{II} =& 2\pi \displaystyle\sum\limits_{i=1}^{N_R} r_i \big( \mathcal{A}_{\rm SP}\left(r_i\right) + \mathcal{A}_{\rm LS}\left(r_i\right) \big) \Delta r,\\
\mathrm{subject~to:}~&0 \le \mathcal{A}_{\rm SP}\left(r_i\right) \le 1,~\mathrm{for}~0 \le r_i \le \frac{1}{2},~\mathrm{and}\\
~&0 \le \mathcal{A}_{\rm LS}\left(r_i\right) \le 1,~\mathrm{for}~0 \le r_i \le \frac{1}{2}.
\end{aligned}
\end{equation}

\noindent Note that in the case where the Lyot stop is a free variable array, then the function being constrained by the optimizer ($\Psi_D$) is no longer a linear function of the free variables. That is because each point in the final field is now determined by products of transmission values in the apodizer and Lyot stop. Although some solvers, such as LOQO, are flexible in accepting non-linear, non-convex programs, convergence on a solution is not guaranteed.

\subsubsection{Focal diaphragm}

For the configurations where the focal plane mask is a diaphragm rather than a spot, our computational approach is distinct. Now, the transmitted region between radii $\rho_0$ and $\rho_1$ is the part of the field computed in the first focal plane, which is in turn directly propagated to the Lyot plane:

\begin{equation}
\begin{aligned}
\Psi_{B}\left(\xi_j, \gamma_k\right) =& 2\pi/\gamma_k \sum\limits_{i=1}^{N_R} r_{i} \mathcal{A}_{\rm SP} \left(r_i\right) J_0\left(2\pi \xi_j r_i /\gamma_k \right) \Delta r, \\
\Psi_C \left(r_i, \gamma_k\right) =& 2\pi/\gamma_k \sum\limits_{j=M_{\rho_0}}^{M_{\rho_1}} \xi_j J_0 \left(2\pi r_i \xi_j / \gamma_k\right) \Psi_{B}\left(\xi_j, \gamma_k\right) \Delta \xi, \\
\Psi_D\left(\zeta_j, \gamma_k \right) =& 2\pi/\gamma_k \sum\limits_{i=1}^{N_r} r_j \mathcal{A}_{\rm{LS}}\left(r_i\right) \Psi_C\left(r_i, \gamma_k\right) J_0\left(2\pi \zeta_j r_i /\gamma_k\right) \Delta r. \\
\end{aligned}
\end{equation}

\noindent Above, $M_{\rho_0}$ and $M_{\rho_1}$ correspond respectively to the lowest and highest integers $j$ satisfying $\rho_0 \le \xi_j \le \rho_1$, where $\xi_j = \left(j - \frac{1}{2}\right)\Delta\xi$. The definitions for the optimization objective and constraints given for the spot FPM configuration, Equations~\ref{eqn:obj_config1a},~\ref{eqn:obj_config2a}, and~\ref{eqn:obj_config3a}, remain valid for the corresponding Configs.\ Ib, IIb, and IIIb, respectively.

\subsection{WFIRST-AFTA SPLC}
\label{subsec:afta_appendix}

The same approach we used to define a discrete, algebraic propagation model for the clear circular aperture SPLC can be applied to an arbitrary telescope aperture. However, the propagation now relies on two-dimensional Fourier transforms rather than Hankel transforms, resulting in a combination of real and imaginary scalar field components. We again code the linear program in AMPL. However, we use the Gurobi\cite{gurobi} package to implement the solver algorithm instead of LOQO, since it better accommodates the much larger size of the two-dimensional problem.

At each stage, we expand the complex exponential of the discrete Fourier transform into cosine and sine terms; doing so reveals simplifications arising from the geometric symmetry of the telescope pupil, thereby reducing the computational complexity and speeding up the optimization. We align the telescope pupil (Figure~\ref{fig:afta_pupil}) so that one of its three symmetry axes coincides with the vertical axis ($y$) in our Cartesian representation. In the first stage of the propagation, this enables us to restrict the bounds of the horizontal Riemann sum to one half of the pupil plane, and also to drop sine terms with a horizontal dependence. The field in the first focal plane is then

\begin{equation}
\begin{aligned}
\Psi_{B_{\rm{Re}}}\left(\xi_u,\eta_v, \gamma_k \right) =&~2/\gamma_k \displaystyle\sum\limits_{j=-N_y}^{N_y} \cos\left(2\pi \eta_v y_j / \gamma_k\right) \displaystyle\sum\limits_{i=1}^{N_x} \mathcal{A}_{\rm SP}\left(x_i,y_j\right) \cos\left(2\pi \xi_u x_i / \gamma_k\right) \Delta x \Delta y, \\ 
\Psi_{B_{\rm{Im}}}\left(\xi_u,\eta_v, \gamma_k \right) =&~2/\gamma_k \displaystyle\sum\limits_{j=-N_y}^{N_y} \sin\left(2\pi \eta_v y_j / \gamma_k\right) \displaystyle\sum\limits_{i=1}^{N_x} \mathcal{A}_{\rm SP}\left(x_i,y_j\right) \cos\left(2\pi \xi_u x_i / \gamma_k\right) \Delta x \Delta y. \\ 
\end{aligned}
\label{eqn:afta_prop_PsiB}
\end{equation}

\noindent The real and imaginary components of the field are distinguished with ``Re'' and ``Im'' subscripts. Combining the facts that $\mathcal{A}_{\rm{SP}}$ is real and symmetric about the vertical axis, it can be shown that (i) $\Psi_{B_{\rm{Re}}}\left(\xi_u,\eta_v, \gamma_k \right)$ has even symmetry over $\xi$ and $\eta$, and that (ii) $\Psi_{B_{\rm{Im}}}\left(\xi_u,\eta_v, \gamma_k \right)$ has even symmetry over $\xi$ and odd symmetry over $\eta$. Using these symmetry properties, we need only evaluate the Riemann sums for $\Psi_{B_{\rm{Re}}}$ and $\Psi_{B_{\rm{Im}}}$ in one quadrant of the focal plane to determine the full field.

As before, the horizontal and vertical coordinates are normalized to the re-imaged telescope pupil diameter. The fabrication process for the WFIRST-AFTA shaped pupils assumes a binary mask array 1000 pixels in diameter\cite{Bala2015JATIS}. Therefore, in order to optimize testbed-ready designs, as in the case of the characterization design presented in Section~\ref{subsec:afta_char_splc}, in Equation~\ref{eqn:afta_prop_PsiB} we set $N_x$ and $N_y$ to 500, and $\Delta x$ and $\Delta y$ to $1/1000$. However, for efficient parameter surveys the spatial resolution can be much coarser, for example $\Delta x = 1/256$.

\subsubsection{Focal occulting spot}

Similar to the circular SPLC, the region within the quadrant where we evaluate $\Psi_B$ depends on the FPM configuration. For the occulting spot FPM, the field is evaluated only in the interior of the occulting spot, since Babinet's principle applies conveniently again when propagating to the Lyot plane.

We represent the discretized profile of the FPM explicitly by the variable array $M\left(\xi_u,\eta_v\right)$. Consistent with the convention used in Equations~\ref{eqn:circ_fpm_def} and~\ref{eqn:aplc_Psi_C}, $M\left(\xi_u,\eta_v\right)$ is the compliment of the mask transmission: zero-valued in the transmitted region and unity in the occulted region. As is necessary in order to approximate round and diagonal features on a Cartesian grid, $M\left(\xi_u,\eta_v\right)$ takes on ``gray'' values between 0 and 1 at the edges of features in the mask profile, in proportion to the fraction of area occulted on the mask array pixel.

If we the sample the interior of the occulting spot of radius $\rho_0$ with $N_{\xi_0}$ horizontal samples at interval $\Delta\xi$ and $N_{\eta_0}$ vertical samples at interval $\Delta\eta$, then the on-axis field propagation to the Lyot plane and final focal plane are modeled as follows:

\begin{equation*}
\begin{aligned}
\Psi_{C}\left(x_i, y_j, \gamma_k \right) = \mathcal{A}_{\rm SP}\left(x_i, y_j\right) -&~\scriptstyle 4/\gamma_k \sum\limits_{v=1}^{N_{\eta_0}} \cos\left(2\pi\eta_v y_i / \gamma_k \right) \sum\limits_{u=1}^{N_{\xi_0}} M\left(\xi_u,\eta_v\right) \Psi_{B_{\rm{Re}}}\left(\xi_u,\eta_v, \gamma_k \right) \cos\left(2\pi\xi_u x_i / \gamma_k \right) \Delta \xi \Delta \eta \\
+&~\scriptstyle 4/\gamma_k \sum\limits_{v=1}^{N_{\eta_0}} \sin\left(2\pi\eta_v y_i / \gamma_k \right) \sum\limits_{u=1}^{N_{\xi_0}} M\left(\xi_u,\eta_v\right) \Psi_{B_{\rm{Im}}}\left(\xi_u,\eta_v, \gamma_k \right) \cos\left(2\pi\xi_u x_i / \gamma_k \right) \Delta \xi \Delta \eta, \\
\end{aligned}
\end{equation*}
\begin{equation}
\begin{aligned}
\Psi_{D_{\rm{Re}}}\left(\zeta_u, \mu_v,\gamma_k\right) =~ \scriptstyle 2/\gamma_k \sum\limits_{j=-N_y}^{N_y} \cos\left(2\pi \mu_v y_j / \gamma_k\right) \sum\limits_{i=1}^{N_x} \mathcal{A}_{\rm LS}\left(x_i, y_j\right) \Psi_C\left(x_i, y_j,\gamma_k\right) \cos\left(2\pi \zeta_u x_i / \gamma_k\right) \Delta x \Delta y,& \\
\Psi_{D_{\rm{Im}}}\left(\zeta_u, \mu_v,\gamma_k\right) =~ \scriptstyle 2/\gamma_k \sum\limits_{j=-N_y}^{N_y} \sin\left(2\pi \mu_v y_j / \gamma_k\right) \sum\limits_{i=1}^{N_x} \mathcal{A}_{\rm LS}\left(x_i, y_j\right) \Psi_C\left(x_i, y_j,\gamma_k\right) \cos\left(2\pi \zeta_u x_i / \gamma_k\right) \Delta x \Delta y.& \\
\end{aligned}
\label{eqn:afta_prop_spotFPM}
\end{equation}

\noindent The field in the Lyot plane, $\Psi_C$, is real and symmetric about the vertical axis. The final focal plane field, $\Psi_D$, retains the same symmetry properties as $\Psi_B$, so again it is most efficient to only evaluate one quadrant.

In our investigations of WFIRST-AFTA solutions, we found that the spatial sampling in the first focal plane is especially critical for maintaining the accuracy of designs with a small inner working angle. When the FPM has an inner radius below $3~\lambda_0/D$, a resolution of $\Delta\xi=\Delta\eta=\frac{1}{8}$ is needed to ensure agreement with high-resolution evaluations of the solution.

Like the circular SPLC, we use the central peak in the first focal plane as a proxy for the star's flux:

\begin{equation}
\Psi_{B_{\rm{Peak}}}\left(\gamma_k\right) = 2/\gamma_k \displaystyle\sum\limits_{j=-N_y}^{N_y} \displaystyle\sum\limits_{i=1}^{N_x} \mathcal{A}_{\rm SP}\left(x_i,y_j\right) \Delta x \Delta y.
\end{equation}

\noindent Then the optimization objective and constraints are defined as follows:

\begin{equation}
\begin{aligned}
\mathrm{Maximize}~\mathcal{O}_I = 2 \sum\limits_{j=-N_y}^{N_y} \displaystyle\sum\limits_{i=1}^{N_x} \mathcal{A}_{\rm SP}\left(x_i,y_j\right) \Delta x \Delta y,\\
\mathrm{subject~to:}~0 \le \mathcal{A}_{\rm SP}\left(x_i,y_j\right) \le \mathcal{A}_{\rm TEL},~\mathrm{for}~0 \le x_i \le \frac{1}{2}, -\frac{1}{2} \le y_j \le \frac{1}{2}, ~\mathrm{and}\\
-10^{-c/2}/\sqrt{2} \le~ \Psi_{D_{\rm Re}}\left(\zeta_u,\mu_v,\gamma_k\right)/\Psi_{B_{\rm Peak}}\left(\gamma_k\right) \le 10^{-c/2}/\sqrt{2},\\
-10^{-c/2}/\sqrt{2} \le~ \Psi_{D_{\rm Im}}\left(\zeta_u,\mu_v,\gamma_k\right)/\Psi_{B_{\rm Peak}}\left(\gamma_k\right) \le 10^{-c/2}/\sqrt{2},\\
\mbox{for}~ \rho_0 \le \sqrt{\zeta_u^2 + \mu_v^2} \le\ \rho_1,\mathrm{~and~}1-w/2 \le \gamma_k \le 1+w/2.
\label{eqn:constraints}
\end{aligned}
\end{equation}

The variable array $\mathcal{A}_{\rm TEL}$ represents the transmission of the telescope pupil, including its central obstruction and support struts, as illustrated in Figure~\ref{fig:afta_pupil}. This condition forces all points in the pupil already obstructed by the telescope to remain opaque in the shaped pupil apodizer solution. When defining $\mathcal{A}_{\rm TEL}$, we pad the telescope obstruction features by 0.25\% of the pupil diameter in order to allow for some alignment error between the shaped pupil apodizer and the relay optics.

\subsubsection{Focal diaphragm}

In the propagation model for the diaphragm FPM configuration, the region of the first focal plane quadrant with non-zero transmission is the only one we compute. In the case of the characterization SPLC design for WFIRST-AFTA described in Section~\ref{subsec:afta_char_splc} this region is bowtie-shaped; for other designs it can be annular. For convenience, we define a FPM variable that is the complement of $M$: $\tilde{M}\left(\xi_u,\eta_v\right)=1-M\left(\xi_u,\eta_v\right)$. Therefore, $\tilde{M}$ is equal to unity in the transmitted region and zero-valued in the occulted region. Starting from Equation~\ref{eqn:afta_prop_PsiB}. The on-axis field propagation to the Lyot plane and final focal plane are modeled as follows:

\begin{equation}
\begin{aligned}
\Psi_{C}\left(x_i, y_j, \gamma_k \right) = &~\scriptstyle 4/\gamma_k \sum\limits_{v=1}^{N_{\eta_1}} \cos\left(2\pi\eta_v y_i / \gamma_k \right) \sum\limits_{u=1}^{N_{\xi_1}} \tilde{M}\left(\xi_u,\eta_v\right) \Psi_{B_{\rm{Re}}}\left(\xi_u,\eta_v, \gamma_k \right) \cos\left(2\pi\xi_u x_i / \gamma_k \right) \Delta \xi \Delta \eta \\
-&~\scriptstyle 4/\gamma_k \sum\limits_{v=1}^{N_{\eta_1}} \sin\left(2\pi\eta_v y_i / \gamma_k \right) \sum\limits_{u=1}^{N_{\xi_1}} \tilde{M}\left(\xi_u,\eta_v\right) \Psi_{B_{\rm{Im}}}\left(\xi_u,\eta_v, \gamma_k \right) \cos\left(2\pi\xi_u x_i / \gamma_k \right) \Delta \xi \Delta \eta, \\
\Psi_{D_{\rm{Re}}}\left(\zeta_u, \mu_v,\gamma_k\right) = &~ \scriptstyle 2/\gamma_k \sum\limits_{j=-N_y}^{N_y} \cos\left(2\pi \mu_v y_j / \gamma_k\right) \sum\limits_{i=1}^{N_x} \mathcal{A}_{\rm LS}\left(x_i, y_j\right) \Psi_C\left(x_i, y_j,\gamma_k\right) \cos\left(2\pi \zeta_u x_i / \gamma_k\right) \Delta x \Delta y,& \\
\Psi_{D_{\rm{Im}}}\left(\zeta_u, \mu_v,\gamma_k\right) = &~ \scriptstyle 2/\gamma_k \sum\limits_{j=-N_y}^{N_y} \sin\left(2\pi \mu_v y_j / \gamma_k\right) \sum\limits_{i=1}^{N_x} \mathcal{A}_{\rm LS}\left(x_i, y_j\right) \Psi_C\left(x_i, y_j,\gamma_k\right) \cos\left(2\pi \zeta_u x_i / \gamma_k\right) \Delta x \Delta y.& \\
\end{aligned}
\end{equation}

The optimization constraints are defined in a manner identical to the previous configuration, except that the points where the contrast is constrained in the image need to be matched to the profile of the focal plane mask, rather than an assumed annular region:

\begin{equation}
\begin{aligned}
-10^{-c/2}/\sqrt{2} \le~ \Psi_{D_{\rm Re}}\left(\zeta_u,\mu_v,\gamma_k\right)/\Psi_{B_{\rm Peak}}\left(\gamma_k\right) \le 10^{-c/2}/\sqrt{2},\\
-10^{-c/2}/\sqrt{2} \le~ \Psi_{D_{\rm Im}}\left(\zeta_u,\mu_v,\gamma_k\right)/\Psi_{B_{\rm Peak}}\left(\gamma_k\right) \le 10^{-c/2}/\sqrt{2},\\
\mbox{for}~ \left(\zeta_u,\mu_v\right)~\mbox{such~that}~\tilde{M}(\zeta_u,\mu_v) > 0,\mathrm{~and~}1-w/2 \le \gamma_k \le 1+w/2.
\label{eqn:afta_constraints_diaphragm}
\end{aligned}
\end{equation}

\noindent For the WFIRST-AFTA designs presented with a 10\% bandwidth ($w=0.10$), we constrained the contrast at 5 wavelengths. For the 18\% bandwidth designs ($w=0.18$), we constrained the contrast at 7 wavelengths.

\acknowledgments
We are grateful to Mamadou N'Diaye and R\'{e}mi Soummer for helpful suggestions and conversations about coronagraph design. We also thank Wesley A. Traub for providing us with his program to estimate the separation-contrast distribution of radial velocity planets. This work was partially funded by NASA Technology Development for Exoplanet Missions (TDEM) grant number NNX09AB96G and by the Jet Propulsion Laboratory of the California Institute of Technology. A. J. E. Riggs is supported by NASA grant number NNX14AM06H, a NASA Space Technology Research Fellowship. Robert J. Vanderbei acknowledges research support from the Office of Naval Research (ONR).


\bibliography{Zimmerman_-_SPLC_JATIS-rev1_arxiv}   
\bibliographystyle{spiejour}   


\vspace{2ex}
\noindent{\bf Neil T. Zimmerman} recently joined the staff of the Space Telescope Science Institute, after postdoctoral research appointments at the Max Planck Institute for Astronomy and at Princeton University. He earned his Ph.D. in astronomy from Columbia University in 2011. Dr. Zimmerman is a member of the SPIE and the American Astronomical Society, and specializes in exoplanet imaging instrumentation and techniques.

\vspace{1ex}
\noindent{\bf A.J. Eldorado Riggs} is a Ph.D. candidate in the Department of Mechanical and Aerospace Engineering at Princeton University. He received a bachelor's degree in physics and mechanical engineering from Yale University in 2011. His research interests include wavefront estimation and control and coronagraph design for exoplanet imaging.

\vspace{1ex}
\noindent{\bf N. Jeremy Kasdin} is a Professor of Mechanical and Aerospace Engineering and Vice Dean of the School of Engineering and Applied Science at Princeton University. He is the Principal Investigator of Princeton's High Contrast Imaging Laboratory.  He received his Ph.D. from Stanford University in 1991. Professor Kasdin's research interests include space systems design, space optics and exoplanet imaging, orbital mechanics, guidance and control of space vehicles, optimal estimation, and stochastic process modeling. He is an Associate Fellow of the American Institute of Aeronautics and Astronautics and member of the American Astronomical Society and the SPIE.

\vspace{1ex}
\noindent {\bf Alexis Carlotti} is an Assistant Astronomer at the Institute of Planetology and Astrophysics in Grenoble, France. He specializes in coronagraph design and adaptive optics instrumentation.

\vspace{1ex}
\noindent{\bf Robert J. Vanderbei} is a Professor in the Department of Operations Research and Financial Engineering at Princeton University. He has degrees in Chemistry (BS), Operations Research and Statistics (MS), and Applied Mathematics (MS, PhD). He is a fellow of the American Mathematic Society, the Society for Applied and Industrial Mathematics, and the Institute for Operations Research and the Management Sciences.

\listoffigures
\listoftables

\end{spacing}
\end{document}